\documentclass[12pt,aps,color,,floatfix]{revtex4}
\usepackage{epsfig}
\usepackage{graphicx}
\usepackage{graphics}

\newcommand{\NPA}{Nucl.\ Phys.\ A}
\newcommand{\PLB}{Phys.\ Lett.\ B}

\newcommand{\PRL}{Phys.\ Rev.\ Lett.}

\newcommand{\PRC}{Phys.\ Rev.\ C}

\begin{document}
\title{Analysis of Dilepton Invariant Mass Spectrum in C+C at 2 and 1 AGeV}
\author{M. Thom\`ere${^1}$, C. Hartnack${^1}$, G. Wolf ${^2}$, J. Aichelin${^1}$}
\affiliation{
$^1$ SUBATECH \\
Laboratoire de Physique Subatomique et des
Technologies Associ\'ees \\
Universit\'e de Nantes - IN2P3/CNRS - Ecole des Mines de Nantes \\
4 rue Alfred Kastler, F-44072 Nantes, Cedex 03, France\\
$^2$ KFKI, P.O. Box 49, H-1525 Budapest Hungary}

\begin{abstract}
Recently the HADES collaboration has published the invariant mass
spectrum of $e^+e^-$ pairs, dN/dM$_{e^+e^-}$, produced in C+C
collisions at 2 AGeV. Using electromagnetic probes, one hopes to get
in this experiment information on hadron properties at high density
and temperature. Simulations show that firm conclusions on possible
in-medium modifications of meson properties will only be possible
when the elementary meson production cross sections, especially in
the pn channel, as well as production cross sections of baryonic
resonances are better known. Presently one can conclude that a)
simulations overpredict by far the cross section at $M_{e^+e^-}
\approx M_{\omega}{^0}$ if free production cross sections are  used
and that b) the upper limit of the $\eta$ decay into  $e^+e^-$ is
smaller than the present upper limit of the Particle Data Group.
This is the result of simulations using the Isospin Quantum
Molecular Dynamics (IQMD) approach.
\end{abstract}
\date{\today}
\maketitle

\section{Introduction}
Theory predicts since long that the properties of hadrons change if they
are surrounded by matter. For baryons this change has been verified in
$\gamma$A reactions where the total photon absorption cross section
\cite{gama} shows a nontrivial dependence on the mass of the target nucleus.
This nontrivial dependence has been interpreted as a change of the properties
of the nuclear resonances in matter \cite{gamth}. It is, however, difficult
to asses whether the observed in-medium modifications have to be attributed
to a change of the resonance properties or to a change of those of
their decay products. Coupled channel calculations provide a mean to answer
this question but presently neither the data are sufficiently precise
nor the theoretical ingredients can be sufficiently well determined in order to
allow for firm conclusions even if for some hadrons like the $\rho$ \cite{rapp} and K \cite{kor}
mesons a lot of progress has been made recently.

The strategy is different for both cases. The study of the strange
mesons takes advantage of the fact that they have to be produced in
a heavy ion reaction, that each strange hadron is accompanied by an
anti-strange one and that the production cross sections are phase
space dominated. Systematic studies of the excitation function and
of the system size dependence of the yields as well as of the
modification of the measured K meson spectra as compared to that 
measured in pp collisions allow for conclusions on the interaction 
of the K's with the environment \cite{kaon}.

The $\rho$ meson can decay into a dilepton pair which - being  an
electromagnetic probe - does not interact anymore with the nuclear
environment. Therefore this dilepton pair carries direct information
on the particle at the time point of its decay in the medium. The
problem is that many resonances and mesons contribute to the
dilepton yield and it is all but easy to determine which particle is
at the origin of the dilepton pair. In order to compare data with
theory, one has to identify all dilepton sources and their
contribution to the dilepton spectra. This superposition of the
different sources is called cocktail plot. If it deviates from
experiment at least one of the sources is not correctly described
and one may start to test how this source is modified by the
hadronic environment.

It was the DLS collaboration which first presented dilepton
invariant mass spectra in heavy ion collisions at beam energies of
around 1 AGeV \cite{dls}. The systematic errors of these exploratory
experiments have been, however, too large to allow for a detailed
conclusion on the behavior of hadrons in matter. Later, at higher
(SPS) energies, the CERES/NA45 collaboration \cite{ceres}  presented
spectra, which were not in agreement with the standard cocktail
plots.  Two theoretical models have been advanced to explain
this difference. Rapp et al. \cite{rapp} calculated the in medium
modification of the spectral function of the $\rho$ in hadronic
matter. With this in-medium change of the spectral function the
theoretical and experimental yields agree. As Eletsky et al. \cite{kapu}  
explained $\rho$ - meson and $\rho$ - baryon interactions compensate each 
other as far as the shift of the pole mass is concerned but collisions 
broaden the width considerably. Gallmeister et al.
\cite{gall} showed on the other side that the discrepancy disappears as well if one
adds to the spectrum the emission of the dileptons from a thermal
$q\bar{q}$ (or hadron-hadron) annihilation using lowest order QCD calculations.

Most recently the NA60 collaboration measured very precisely
the invariant mass spectrum of dileptons in the $\rho$ mass region
\cite{na60} but it is still debated whether the
discrepancy between cocktail plot and data is due to a modification of 
hadronic properties or due to annihilation processes.  Additional information may be obtained 
from the $p_t$ spectra \cite{renk} because each emission source shows a specific 
transverse momentum pattern. However, consensus about the relative importance of
the different possible production mechanism has not been obtained yet.

To clarify this question it is necessary to study the dilepton production at lower energies
where quarks remain bound in hadrons and hadron hadron annihilations are rare.
Then the process
proposed by Gallmeister is absent and thermal production does not play a decisive
role. In addition one has to investigate small systems where direct
collisions dominate over the production in the participant heat bath.

Recently the HADES collaboration has published the dilepton invariant
mass spectrum for the reaction C + C at 2 AGeV \cite{hades}.  This system is small 
and at this energy the formation of a
quark phase is beyond reach as the analysis of many other observables has
shown. It may therefore serve to solve the question of how the $\rho$ meson changes
in a hadronic environment provided that it can be proven that all the other
ingredients of the cocktail plot are well under control.

It is the purpose of this article to investigate in detail the dilepton invariant mass
spectra using one of the presently available programs which simulate heavy ion reactions
on an event by event basis, the Isospin Quantum Molecular dynamics (IQMD)
approach. The main objective is to find out whether the present dilepton data are sufficiently precise
to allow for conclusions on the theoretically predicted change of the particle properties 
in a nuclear environment or to identify the obstacles on the way to achieve this goal.
We concentrate in this exploratory study on the most significant modifications: 
mass shifts and changes of the decay width.    

Before we present the results of our simulations we start out with a short presentation of the 
model and a discussion of our present theoretical and experimental knowledge on all the elementary processes
which contribute to the dilepton spectra and of how they are implemented in our
simulation program.

\section{The IQMD model}
The semi-classical IQMD program \cite{hartn} simulates heavy ion 
reactions on a event by event basis and is one of the standard analyzing tools for heavy ion 
reactions at and below 2 AGeV. In this program hadrons interact by potentials and by collisions.
The former ones are Br\"uckner G-matrix parameterizations for the baryons or parametrized
meson-baryon potentials. Thus nuclei are bound objects with a binding energy
following the Weizs\"acker mass formula. If two hadrons come closer than $r = \sqrt{\sigma_{tot} / \pi}$
they collide. If several exit channels are available a random number determines which one is realized.
The relative weight is given by the relative cross section. The momenta and the mass 
(if the particles have a finite width) of the hadrons in the final state
are randomly determined.
Their distribution follows either experimental measurements or phase space, if experimental results 
are not available. In the standard version \cite{hartn} of the program, nucleons as well as baryonic resonances, pions 
and kaons are the particles which are propagated. 

For the investigation presented here we have
added production cross sections of all particles which
may contribute to the invariant mass spectrum of dileptons: np bremsstrahlung,
$\eta$ Dalitz and direct decay,  $\omega$ (Dalitz and direct) and $\rho$ decay, $\Delta$ Dalitz
decay and $\pi^0$ Dalitz decay. Because we concentrate on a very light system, 
where the probability that mesons have secondary interactions is small, it has not 
been necessary to add the (largely unknown) meson absorption or rescattering cross sections
or to use off-shell transport approaches.
When these particles are produced we use the branching ratios of the Particle Data Group 
\cite{rev} to determine their contribution to the dilepton spectrum.

\section{elementary dilepton cross sections}

\subsection{$\pi^0$ production and decay}
\subsubsection{$\pi^0$ decay into dileptons}
At low invariant mass the overwhelming number of dileptons comes from the decay of
$\pi^0$ mesons which can decay into dileptons via $\pi^0 \to e^+e^-\gamma$.
The shape of the mass distribution of a dilepton in a $\pi^0$ Dalitz decay is given by \cite{kroll}:

\begin{equation}
\frac{dN}{dM}=\frac{1}{M}(1+2\frac{m_{e^-}^2}{M^2})(1-\frac{M^2}{m_{\pi_0}^2})^3\sqrt{1-4\frac{m_{e^-}^2}{M^2}}.
\end{equation}
$m_{\pi_0}$ is the mass of the $\pi_0$, $m_{e^-}$ the electron mass
and M that of the dilepton pair. We take the branching ratio
BR$(\pi^0 \to e^+e^-\gamma)$ as 0.01198.

\subsection{$\eta$ production and decay}
\begin{figure}[!ht]
\epsfig{file=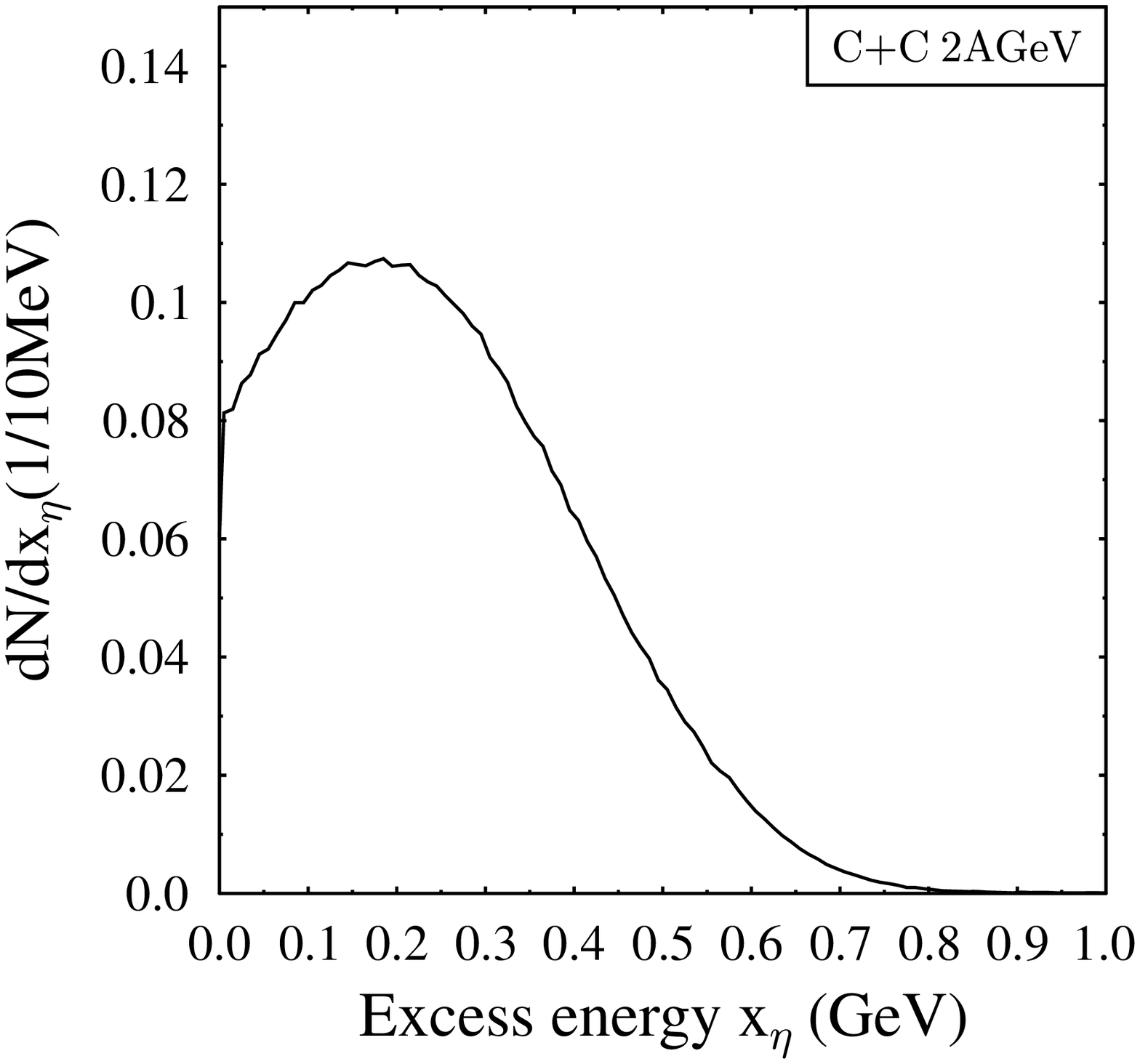,width=10cm} \vspace*{-.3cm}
\epsfig{file=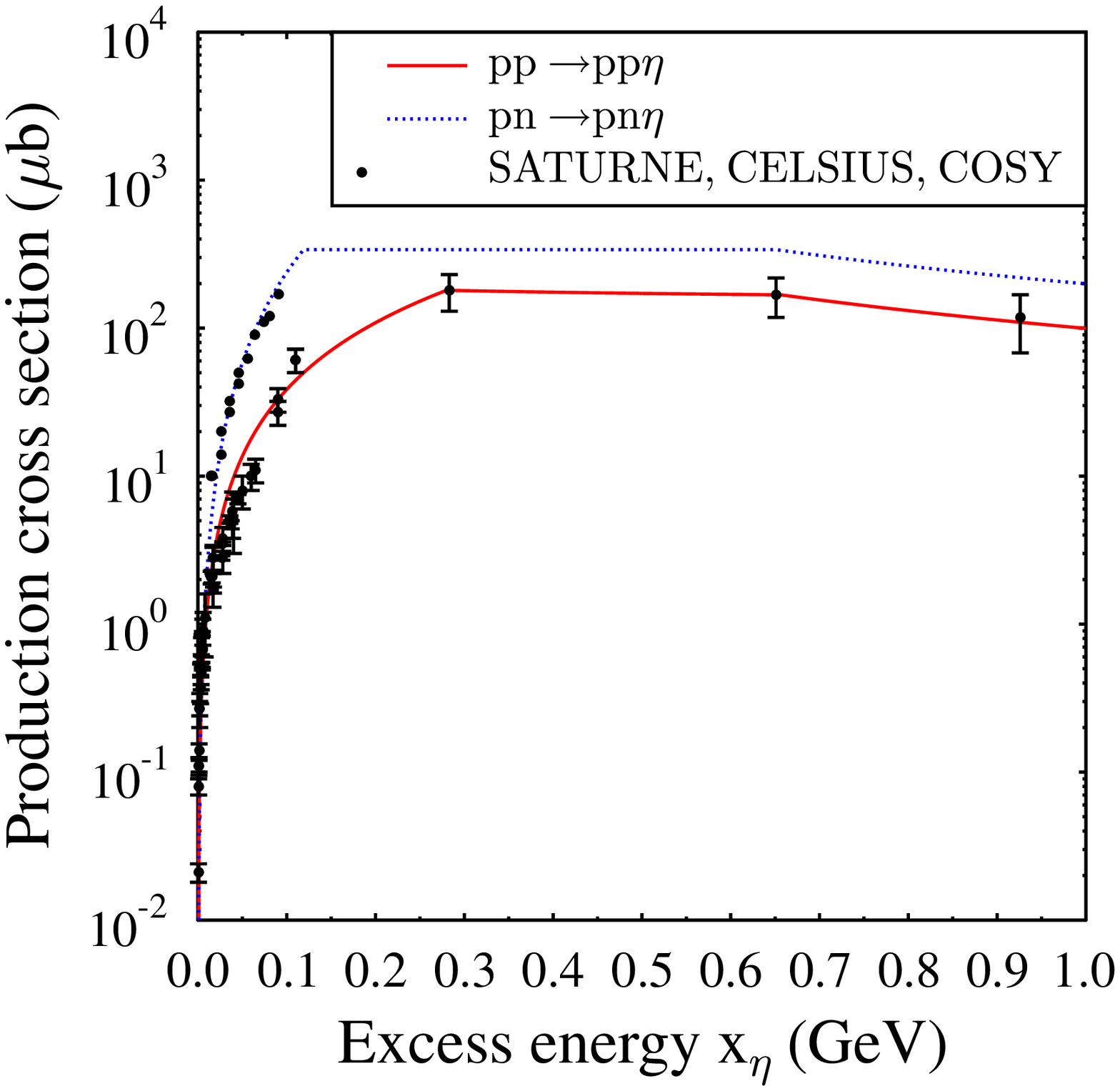,width=10cm} \vspace*{-.3cm}
\caption{(Color online) Top: excess energy, $x_\eta$,  distribution of NN collisions in the reaction 2 AGeV C+C.
Bottom: production cross section of the $\eta$ meson.
The solid curves are fits (eq. \ref{fitppeta}) to the data \cite{csppeta2,donneespneta}}.
\label{csppeta}
\end{figure}

In the energy regime which is of interest here, the $\eta$
production in pp collisions has been well studied by the TAPS
\cite{csppeta2} and the DISTO collaboration \cite{ppn*eta}. This can
be seen in fig. \ref{csppeta} which shows on top the distribution of 
the $\eta$ excess energies in the nucleon nucleon collisions for the reaction 
C+C at 2 AGeV. The excess energy $x_{\eta}$ is defined as
\begin{equation}
x_{\eta}=\sqrt{s}-2M_N-M_{\eta}.
\label{cross}
\end{equation}
We see that excess energies below 0.6 GeV are most relevant for this
reaction. In the bottom part of fig. \ref{csppeta} we display the
world data points for $\eta$ production in elementary NN collisions
\cite{csppeta2, ppn*eta, cernhera84}. Whereas the cross section
$\sigma(pp\to pp\eta)$ is known over the whole excess energy interval
which is relevant for our investigation the $\sigma(pn\to pn\eta)$
cross section is known only up to an excess energy of $x_{\eta} =
0.12$ GeV. Thus we have to extrapolate this cross section into the
relevant excess energy domain. This extrapolation leaves a lot of
freedom even if the $\eta$ meson production cross section has been
measured in heavy ion reactions by the TAPS collaboration. The
reason is that in heavy ion reactions a multitude of processes may
modify the elementary cross section at the same nominal energy.
These processes and the consequences will be discussed later. We
parametrize the $\sigma(pn\to pn\eta)$ and $\sigma(pp\to pp\eta)$ cross
section by a fit using the form
\begin{equation}
\sigma(x_\eta)=ax_\eta^b
\label{fitppeta}
\end{equation}
with $a=1213.8, a=162.1, a=99.6$ $\mu b$ and b=1.50, b=-0.08, b=-1.24 for
excess energies of $x_\eta < 283$ MeV, 283 MeV $< x_\eta <$ 651 MeV , $x_\eta > 651$ MeV for pp
collisions and
$a=25623, a=324.3, a=199 $ $\mu b$ and b=2.03, b=-0.08, b=-1.24 for
excess energies of $x_\eta < 200$ MeV, 200 MeV $< x_\eta <$ 651 MeV , $x_\eta > 651$ MeV for np
collisions assuming that at large excess energies the np cross section is twice the pp cross section.
These fits are also displayed in fig. \ref{csppeta}.
We parametrize the shape of the mass distribution of the $\eta$ by \cite{woehri}
\begin{equation} \label{meta}
{
\frac{dN}{dM}=\frac{(1+2\frac{m_{e^-}^2}{M^2})\sqrt{1-4\frac{m_{e^-}^2}{M^2}}}{(m_{\eta}^2-M^2)^2+
[m_{\eta}(\frac{\Gamma_{\eta} m_\eta}{M}\frac{(M^2/4-m_{e^-}^2)^{3/2}}{(m_{\eta}^2/4-m_{e^-}^2)^{3/2}})]^2}
}
\end{equation}

with $m_{\eta}=0.547$ GeV and $\Gamma_{\eta}$=1.18 keV.

\subsubsection{Contribution of the N*(1535)}
\begin{figure}[!ht]
\epsfig{file=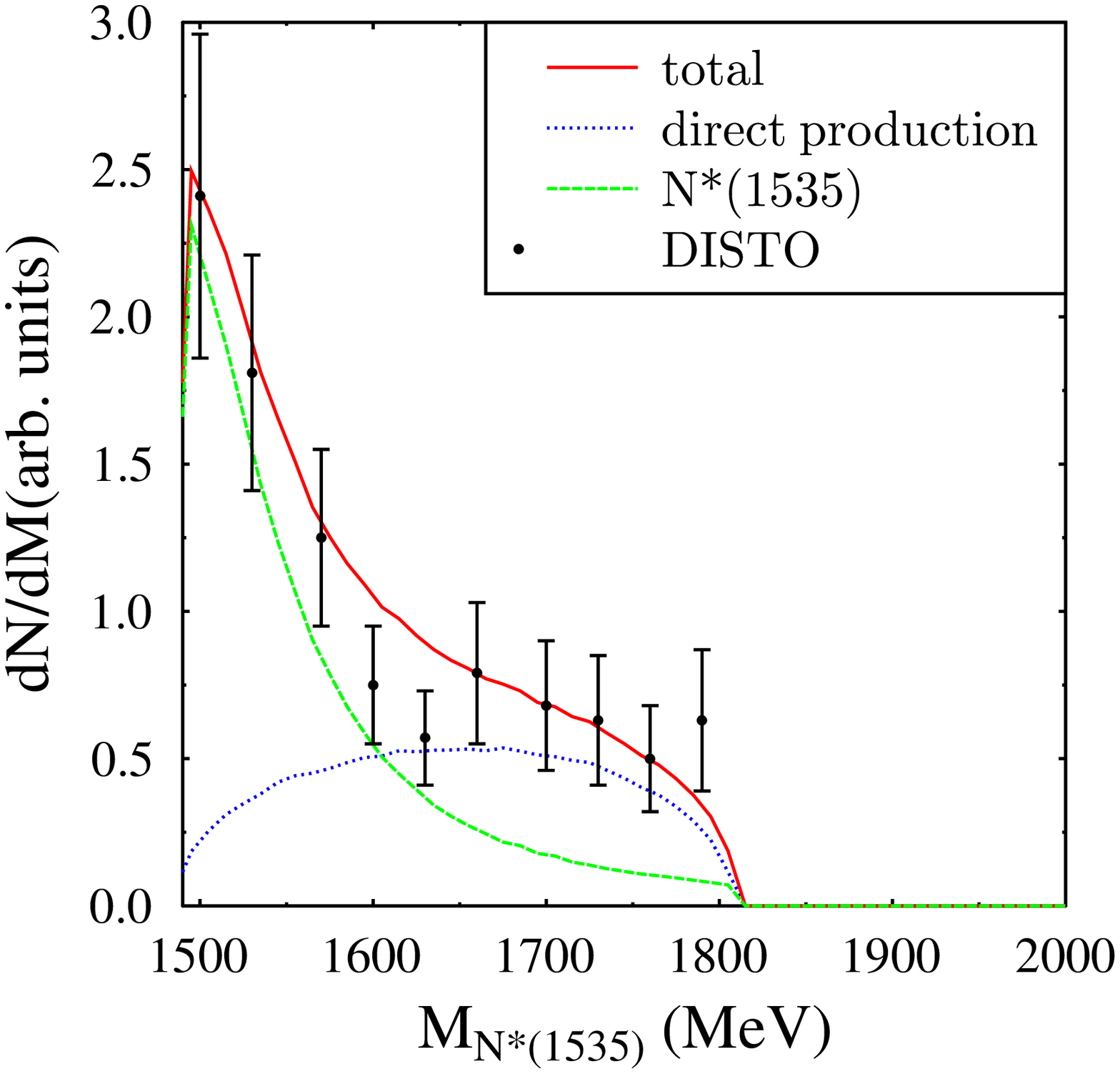,width=7cm} \vspace*{-.3cm}
\epsfig{file=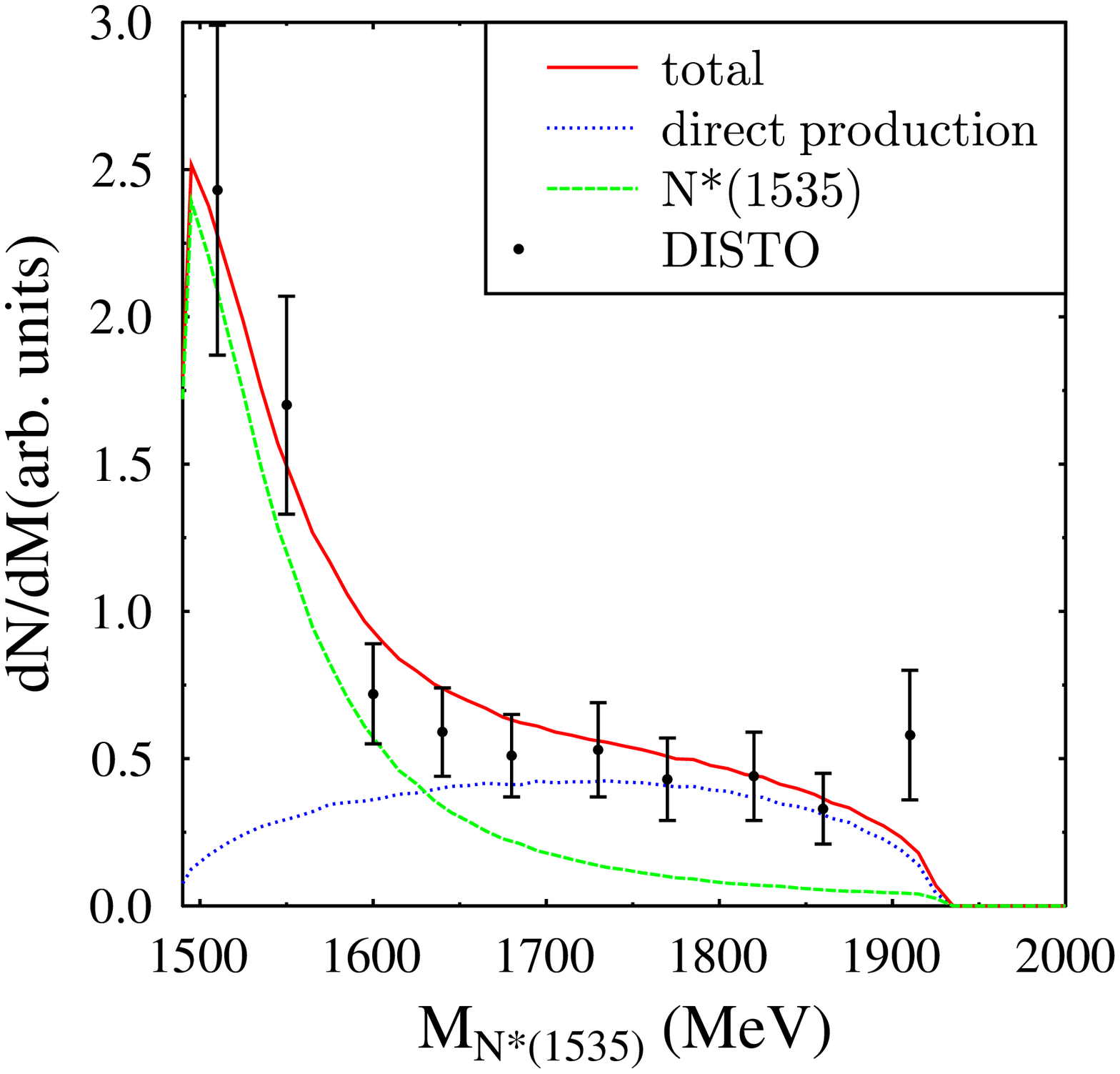,width=7cm} \vspace*{-.3cm}
\epsfig{file=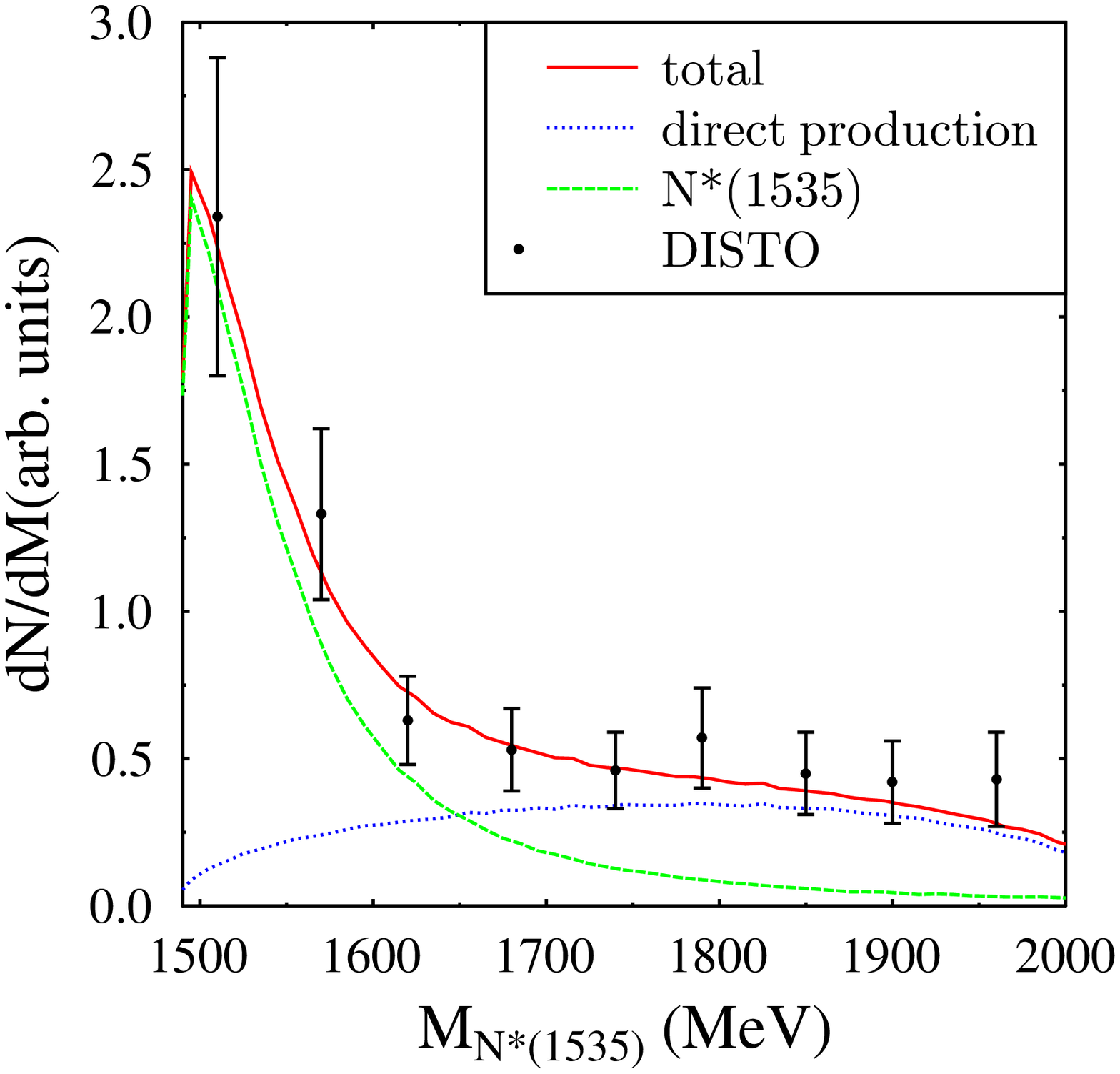,width=7cm} \vspace*{-0.0cm}
\caption{(Color online)  Simulated invariant mass spectrum of the outgoing proton
and the $\eta$ meson in the p p $\to$ p p $\eta$ reaction for three
beam energies (2.15 GeV, 2.5 GeV and 2.85 GeV). The curves
represent the sum of the contributions from the two productions
channels of the $\eta$, the direct production  and that via the
N*(1535) resonance. The data \cite{ppn*eta} have no absolute normalization. We normalize
them here to our result at $M_{N^*(1535)}=1500$ MeV/$c^2 $} \label{mres}
\end{figure}

\begin{figure}[!ht]
\epsfig{file=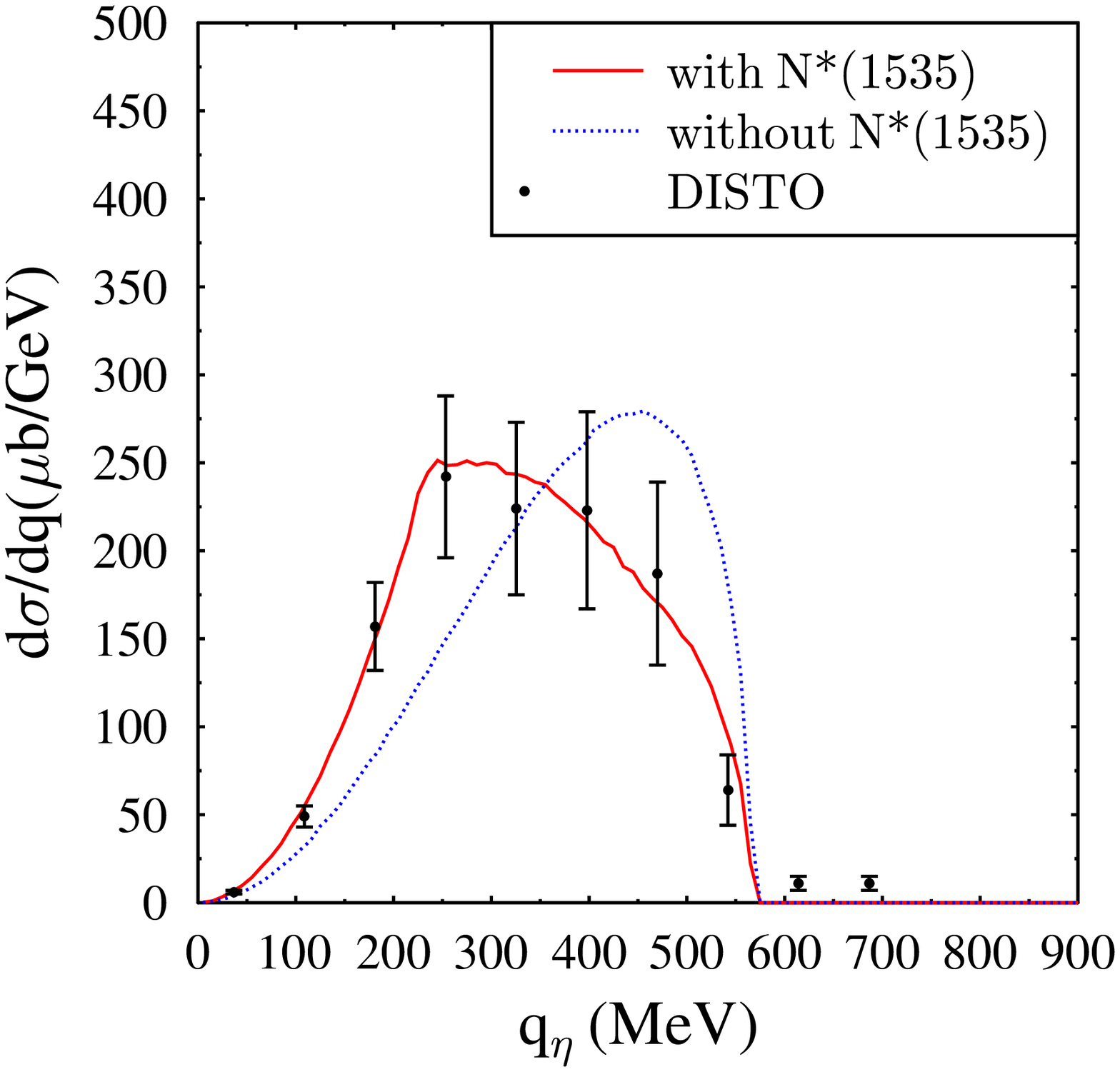,width=6cm} \vspace*{-.2cm}
\epsfig{file=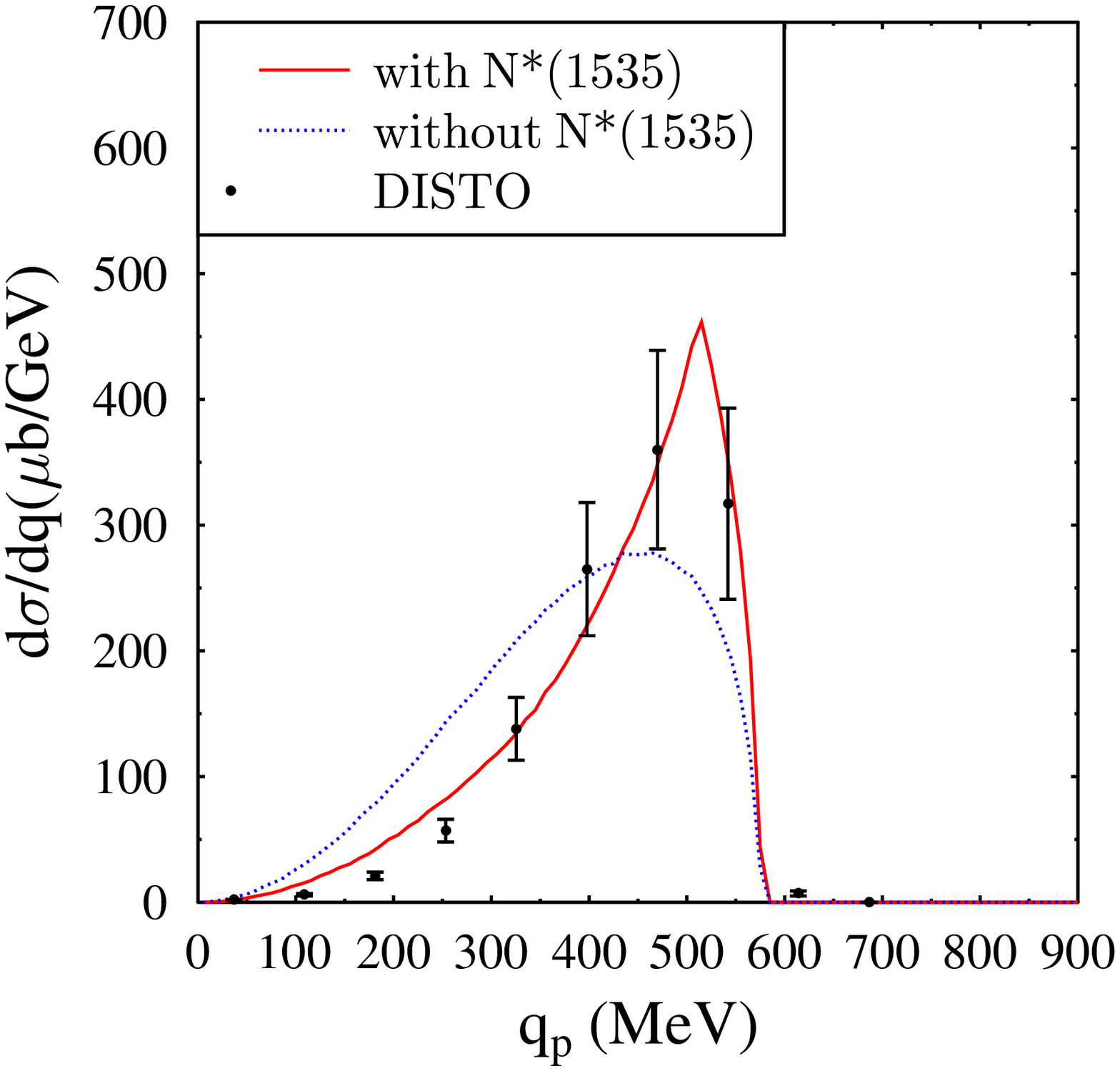,width=6cm} \vspace*{-.2cm}
\epsfig{file=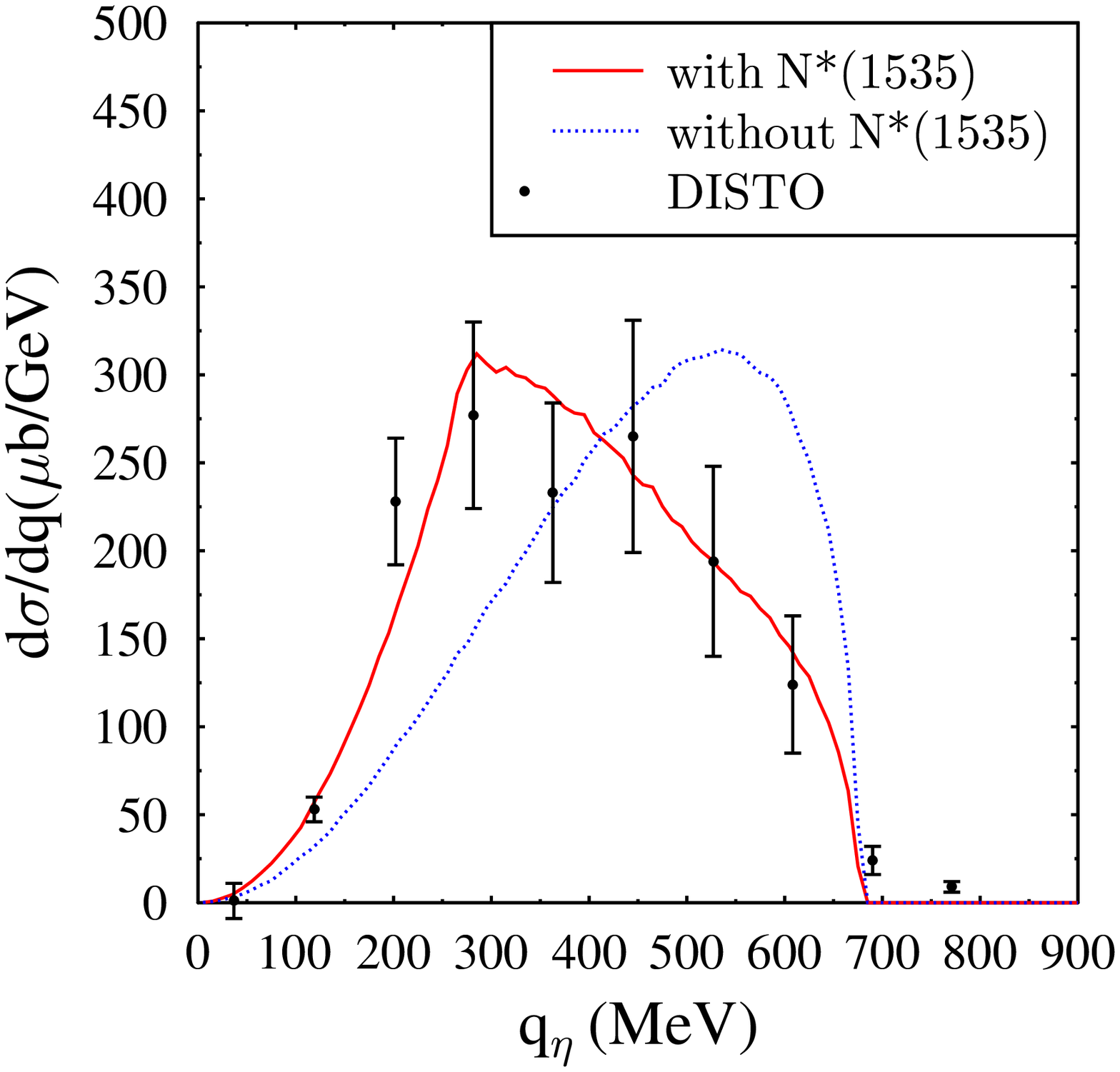,width=6cm} \vspace*{-.2cm}
\epsfig{file=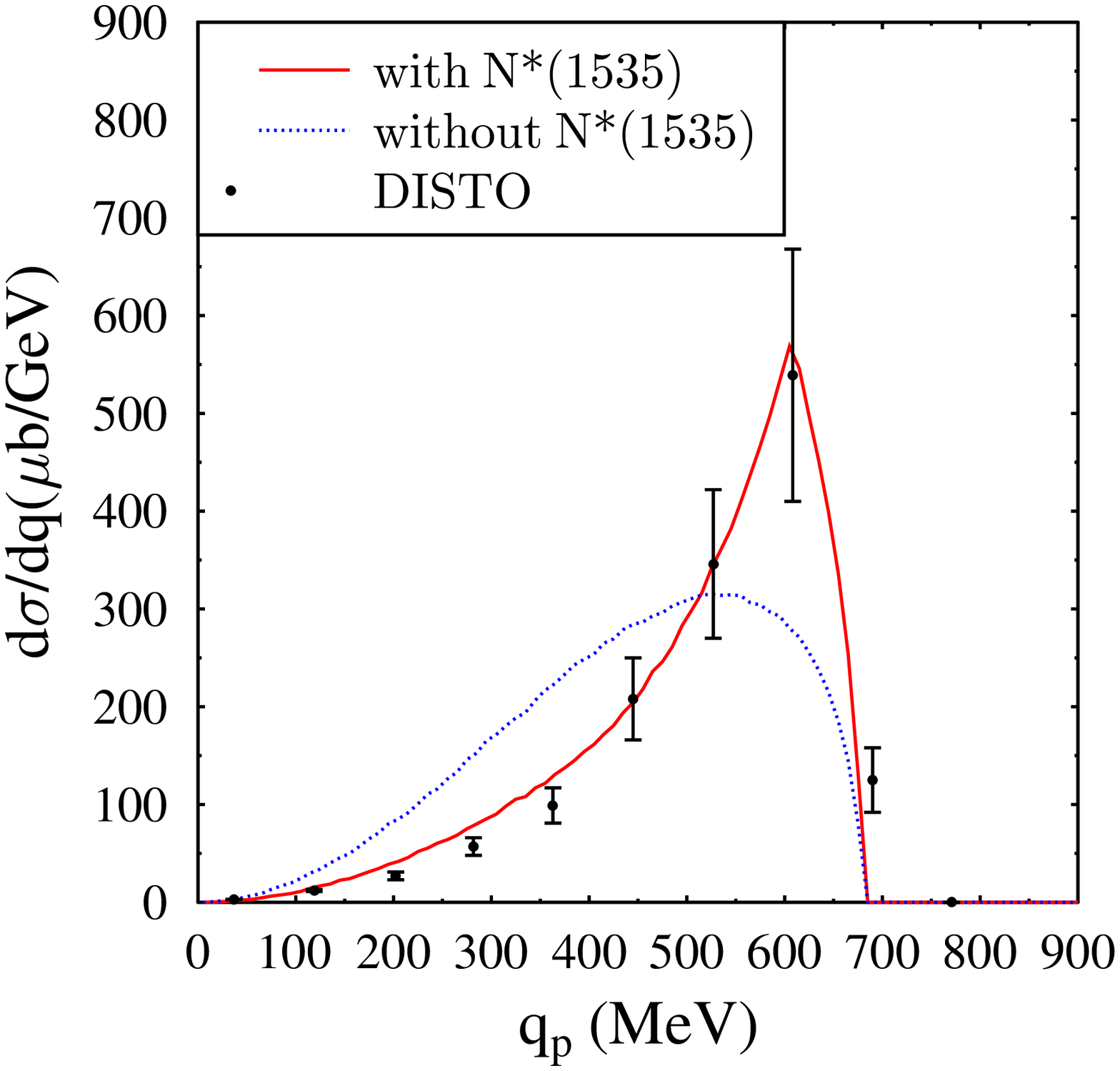,width=6cm} \vspace*{-.2cm}
\epsfig{file=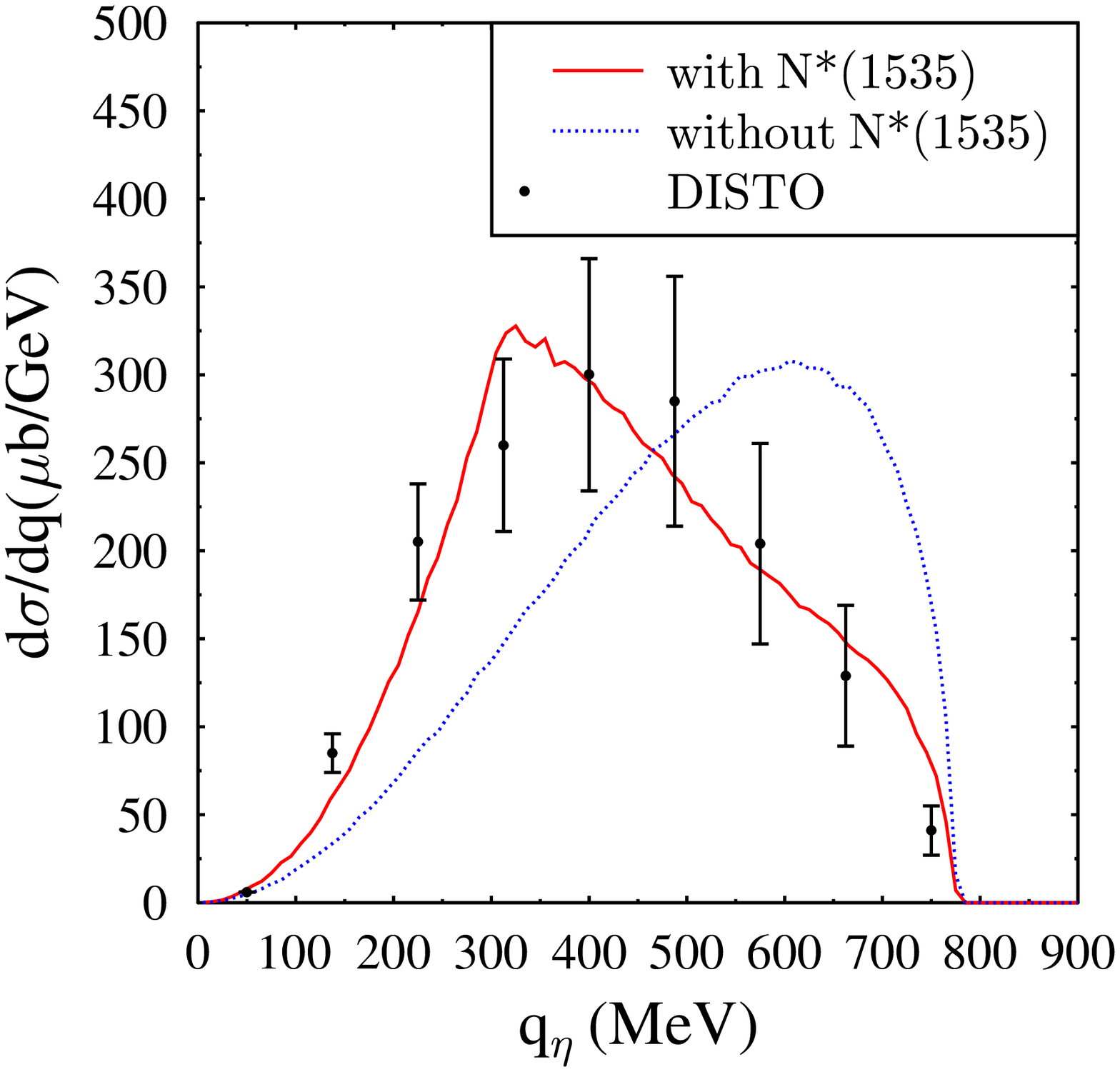,width=6cm} \vspace*{-.2cm}
\epsfig{file=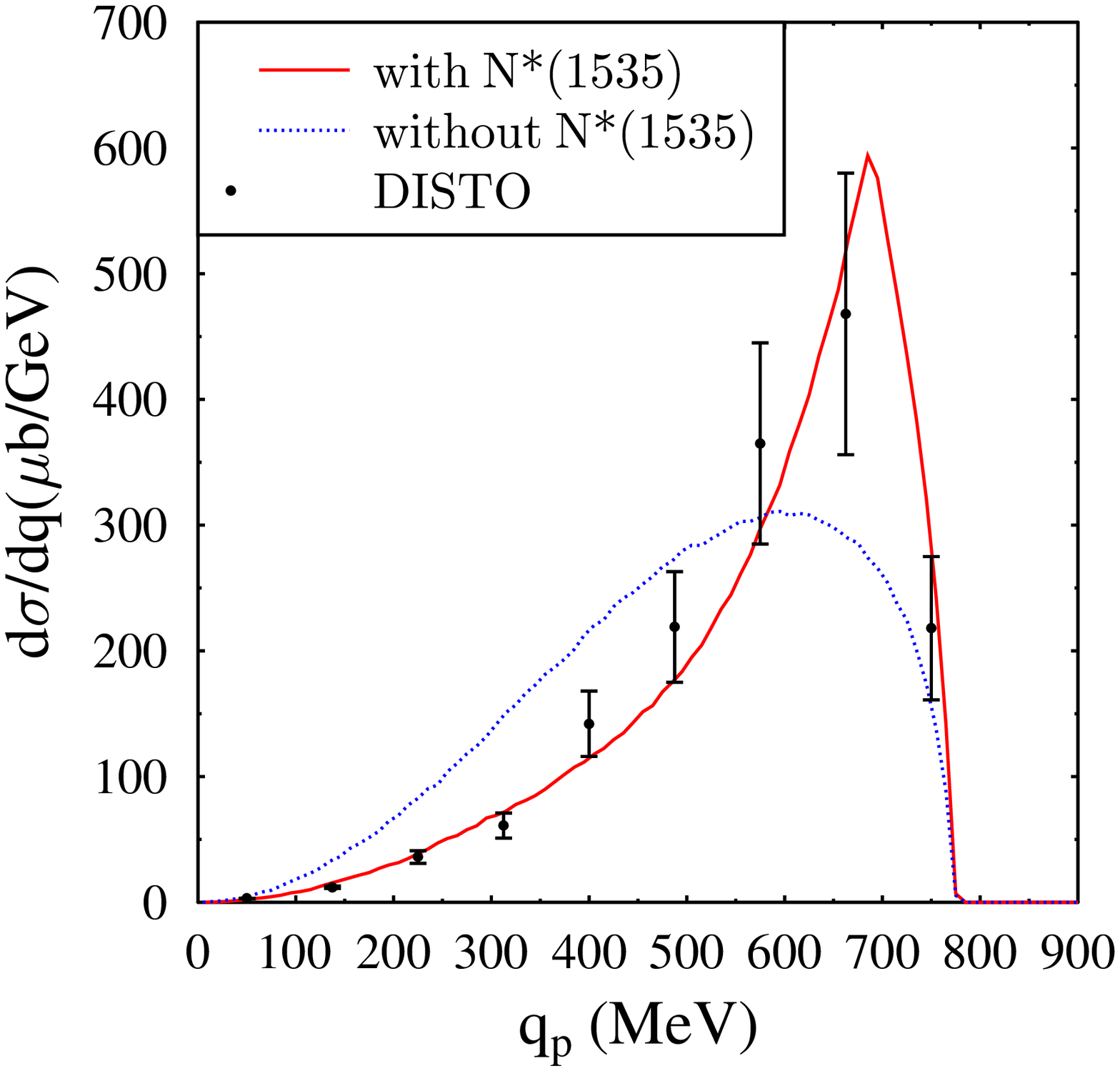,width=6cm} \vspace*{-0.0cm}
\caption{(Color online) Differential cross section in IQMD as a function of the center of mass momentum of the $\eta$ meson (left)
and as a function of the proton momentum in the pp rest system (right) for $pp\to pp\eta$ collisions at
different beam energies, $E_{beam}$ = 2.15 GeV (up), $E_{beam}$ = 2.5 GeV (middle) and
$E_{beam}$ = 2.85 GeV (down). Solid lines represent $\eta$ production including the contribution
of the  $N^*(1535)$ resonance and dashed curves represent the direct production
via an uniform three body phase space distribution. The experimental data are form ref. \cite{ppn*eta} }
\label{impetaproton}
\end{figure}

The very detailed experimental investigation of the $\eta$ production in pp collisions at excess energies
of 324, 412, and 554 MeV (corresponding to beam energies of $E_{beam}$=2.15, 2.5 and 2.85 GeV)
by the DISTO collaboration \cite{ppn*eta} allows to identify the different production channels
by analyzing the p$\eta$ invariant mass spectrum. It turned out, as predicted by theory
\cite{csppetatheo1,csppetatheo2},
 that there are essentially two
channels, a direct production channel and a production via the  $N^*(1535)$ resonance.
The direct contribution follows the three body phase space for the $pp\to pp\eta$ reaction.
The experimental mass  distribution of the  $N^*(1535)$ resonance created in the reaction
p p $\to N^*(1535)$ p can be described by a Breit-Wigner distribution of the form \cite{ppn*eta}:
\begin{equation} \label{massen*}
{
\sigma(M) = \frac{AM_R^2\Gamma_R^2}{(M_R^2-M^2)^2+M_R^2\Gamma_R^2x^2(M,M_R)}
}
\end{equation}
with
\begin{equation} \label{xmassen*}
{
x(M,M_R) = b_\eta\frac{q_\eta(M)}{q_\eta(M_R)}+b_\pi\frac{q_\pi(M)}{q_\pi(M_R)}
}
\end{equation}
where $b_\eta$ is the branching ratio of the decay N*(1535) $\to$ N $\eta$ (which we assume to be 55\%), $b_\pi$ is the branching ratio of the decay N*(1535) $\to$ N $\pi$ (which counts for 45\%).
$q_\pi$ and $q_\eta$ are the momenta of $\pi$ and $\eta$ in the frame of the resonance
and are given by:
\begin{equation} \label{qetamassen*}
{
q_\eta(M_{N^*})=\sqrt{(\frac{M_{N^*}^2-M_p^2+M_\eta^2}{2M_{N^*}})^2-M_\eta^2}
}
\end{equation}
and
\begin{equation} \label{qpimassen*}
{
q_\pi(M_{N^*})=\sqrt{(\frac{M_{N^*}^2-M_p^2+M_\pi^2}{2M_{N^*}})^2-M_\pi^2}
.}
\end{equation}
We note in passing that in reference \cite{ppn*eta} the square on the $x$
in eq. \ref{massen*} has been forgotten.
In fig. \ref{mres} we display for the three energies which have been measured by the DISTO
collaboration \cite{ppn*eta} the total experimental and theoretical $p\eta $ invariant mass distribution
as well as the different contributions to the theoretical curve. The experimental data are best reproduced  for $M_R$ = 1.530 GeV and
$\Gamma_R$ = 150 MeV. As expected, the $N^*(1535)$ resonance enhances the low invariant mass
part as compared to phase space.
How the resonance production modifies the spectra of proton and $\eta$ as compared to the production
according to the three body phase space is shown in fig. \ref{impetaproton}. On the left
hand side we display the center of mass momentum of the $\eta$, on the right hand side the
proton momentum in the pp rest frame. Choosing these variables allows for a comparison with
the experimental results. We see clearly the consequence of the $\eta$ resonance
production and therefore it will be difficult to separate the modification of the
$\eta$ in the medium from that of the N*(1535) resonance. Both will show up as a modification of
the dilepton spectra.

\subsubsection{$\eta$ decay into dileptons}
With a branching ratio of $6 \cdot 10^{-3}$ \cite{rev} the $\eta$ decays into $e^+e^-\gamma$.
The shape of the invariant mass distribution of the dilepton pair is given by \cite{kroll}
\begin{equation}
\frac{dN}{dM}=\frac{1}{M}(1+2\frac{m_{e^-}^2}{M^2})(1-\frac{M^2}{m_{\eta}^2})^3\sqrt{1-4\frac{m_{e^-}^2}{M^2}}.
\end{equation}
$m_{\eta}$ is the mass of the $\eta$, $m_{e^-}$ the electron mass and M that of the dilepton pair.
It has been shown that this expression has to be multiplied with an electromagnetic form factor. With
\begin{equation}
(\frac{dN}{dM})_{tot}=F(M^2)*\frac{dN}{dM}
\end{equation}
where
\begin{equation}
F(M^2)=(\frac{1}{1-\frac{M^2}{\Lambda_\eta^2}})^2
\end{equation}
with $\Lambda_\eta$ = (0.72 $\pm$  0.09) GeV  one finds good
agreement with data \cite{woehri}. In addition to the three body
decay there may  also be a two body one into a dilepton
pair. The Particle Data Group \cite{rev} quotes as an upper limit a
branching ratio of $7.7\cdot 10^{-5}$. We include this value in our
standard calculation (standard will be explained later).

\subsection{$\omega$ production and decay}
\begin{figure}[!ht]
\begin{center}
\includegraphics[width=7.0cm]{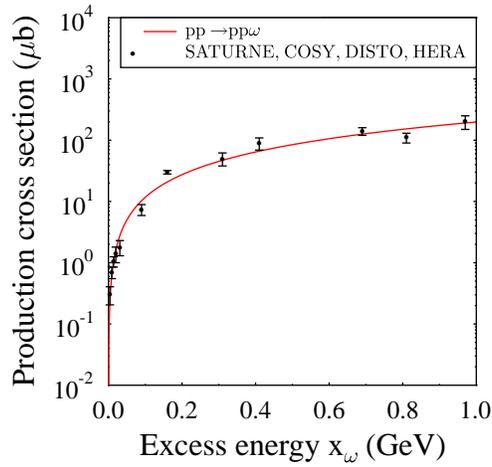}
\end{center}
\caption{(Color online) Production cross section of the  $\omega$ in pp collision up to an
excess energy of 440 MeV and our fit of the form  $\sigma=ax_\omega^b$. The data
are from ref \cite{cos,sat,disto}.}
\label{csppomega}
\end{figure}

The $\omega$ production in pp collisions for excess energies below
440 MeV has been studied at COSY \cite{cos}, at SATURNE \cite{sat}
and by the DISTO \cite{disto} collaboration. The cross section as
well as our fit of the form $a{x_\omega}^b$ where ${x_\omega}$ is the
excess energy in MeV, $a=(192.204\pm 8.622)$ $\mu b$ and
$b=1.12182\pm 0.1077$ is shown in fig. \ref{csppomega}. We include
in our simulation as well the endothermic ($\sqrt{s_0}\simeq$m$_\omega$-m$_\pi$= 643
MeV) reaction $\pi$+N$\to \omega$+N. Because $\pi$'s have usually only
a small energy this reaction is less important than the baryonic channel. 
The experimental
data have been parametrized \cite{pinomegarho} by
\begin{equation}
\sigma_{\pi N\to \omega N} (mb)=\frac{1.38(\sqrt{s}-
\sqrt{s_0})^{1.6}}{0.0011+(\sqrt{s}-\sqrt{s_0})^{1.7}}
\end{equation}
with $\sigma$ in mb, $\sqrt{s}$ and $\sqrt{s_0}$ in GeV.
 $\sqrt{s_0}=m_N+m_{\omega}$ (1.721 GeV for $\omega$ in vacuum) is the threshold energy. It has been suggested that the production of the $\omega$
passes by the excitation of baryon resonances \cite{tsuna,fuchs}
where the $N^*(1535)$ plays  a prominent role having a substantial
branching ratio into the $N\omega$ channel \cite{fae,lutz,post}. It 
produces $\omega$ mesons with masses well below 783 MeV. If this
were the case the strong $\omega$N coupling would lead to a strong
off-shell contribution to $d\sigma / dM$ (M being the invariant mass
of the dilepton pair) at invariant masses well below the free
$\omega$ mass peak. This off-shell $\omega$ production would even
dominate the dilepton spectra up to excess energies of several
hundred MeV. Only very recently calculations of the spectral
function have been advanced which exploit the available $\gamma N$
and $\pi N$ data in a coupled channel analysis \cite{lutz,mos}.

\begin{figure}[!ht]
\begin{center}
\includegraphics[width=7.0cm]{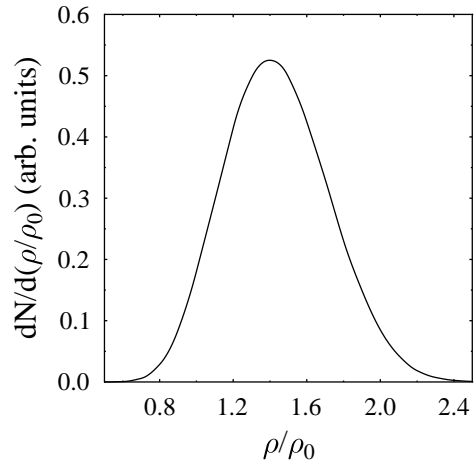}
\end{center}
\caption{(Color online) Distribution of the density at the omega production points
in units of the normal nuclear matter density $\rho_0$ for the
reaction C+C at 2 AGeV.} \label{density}
\end{figure}

A $\gamma A \to \omega$ experiment was recently performed by the CBELSA/TAPS collaboration.
They observed that the pole mass decreases with increasing density of the environment.
\cite{taps}. For momenta less than 500 MeV/$c^2$, they observed an
$\omega$ pole mass of M=[$722^{+2}_{-2}$(stat)$^{+35}_{-5}$(syst)] MeV/c$^2$ for an average density of
0.6 $\rho_0$. Unfortunately, no significant measurement of the width was obtained due to the
dominance of the experimental resolution. Using this data and the Brown-Rho scaling formula:
\begin{equation}
m_{\omega}^*=m_{\omega}^0*(1-\alpha\frac{\rho}{\rho_0})
\label{scal}
\end{equation}
we find $\alpha$=0.13. 
Fig. \ref{density} shows the density distribution at the $\omega$
production points for a C+C collision at 2AGeV. The average density
of $<\rho>$=1.394 $\rho_0$ is twice as large as for the TAPS
experiment. Applying eq. \ref{cross} we obtain a wide distribution
around the average pole mass of M= 641 MeV.

In our simulation we have the option to use this in-medium mass
modification. Because there are no conclusive results on the width
we kept the free value of 8 MeV. The shape of the invariant mass distribution of dileptons from the
$\omega$ decay is given by the Breit-Wigner distribution :

\begin{equation}
\label{momega}
{
\frac{dN}{dM}=\frac{(1+2\frac{m_{e^-}^2}{M^2})
\sqrt{1-4\frac{m_{e^-}^2}{M^2}}}{(m_{\omega}^2-M^2)^2+
[m_{\omega}(\frac{\Gamma_{\omega} m_\omega}{M}
\frac{(M^2/4-m_{e^-}^2)^{3/2}}{(m_{\omega}^2/4-m_{e^-}^2)^{3/2}})]^2}
}
\end{equation}
with $\Gamma_{\omega}$=8 MeV and $m_{\omega}$ as defined in eq. \ref{scal}.


Another uncertainty is the production of the $\omega$ in pn reactions. In meson
exchange models the relative strength of the production in pp and pn reactions
depends strongly on the quantum number of the exchanged mesons. Neglecting
possible differences due to initial and final state interactions, we expect
$\sigma(pn\rightarrow pn\omega)$/$\sigma(pp\rightarrow pp\omega)$ = 5, if only
isovector mesons ($\pi , \rho$) are exchanged \cite{bar}. The two data
points for the reaction $np \rightarrow  d\omega$ point toward an enhancement
of the pn cross section as compared to the pp cross section \cite{bar}. The
error bars are, however, too large in order to quantify this enhancement.
In our simulations we assume $\sigma(pn\rightarrow
pn\omega)=b*\sigma(pp\rightarrow pp\omega)$ with different values of b.

The $\omega$ contributes to the dilepton spectrum in two
different ways. Either it decays directly into a dilepton pair whose
invariant mass equals that of the $\omega$ meson or the dilepton
pair is accompanied by a $\pi_0$ meson. For the latter channel the shape of the
dilepton invariant mass distribution has been parametrized by Kroll
\cite{kroll}
\begin{equation} \label{omegadal}
\frac{dN}{dM}=\frac{1}{M}(1+2\frac{m_{e^-}^2}{M^2})[
(1+\frac{M^2}{m_{\omega}^2-m_{\pi^0}^2})^2-4
\frac{m_{\omega}^2 M^2}{(m_{\omega}^2-m_{\pi^0}^2)^2}]^{3/2}
\sqrt{1-4\frac{m_{e^-}^2}{M^2}}
\end{equation}
with M being the invariant dilepton mass. This Dalitz type decay has
to be corrected by an electromagnetic form factor \cite{woehri}
\begin{equation}
(\frac{dN}{dM})_{tot}=F(M^2)*\frac{dN}{dM}
\end{equation}
with
\begin{equation}
F(M^2)=\frac{a^4}{(a^2-M^2)^2+a^2b^2}
\end{equation}

and a = 0.6519 GeV , b = 0.04198 GeV in order to be in agreement with data.
The branching ratios into the two channels are given
by $5.9\cdot 10^{-4} (7.14\cdot 10^{-5})$ for the $e^+e^-\pi$ $(e^+e^-)$ channel
\cite{rev}.
Both, the unknown pn cross section as well as the little known off-shell
contribution at small excess energies make it difficult to predict the
$\omega$ contribution at invariant dilepton masses between 0.6 and 0.8 GeV.

\subsection{$\rho$ production and decay}
In our simulation the $\rho$ meson can be produced in three channels:
$N N \to N N \rho$, $\pi N \to \rho N$ and $\pi^+ \pi^- \to \rho$.

The few experimental data points of the total
cross section in the $N N \to N N \rho$ channel have been fitted by  \cite{pprho}
\begin{equation}
\sigma_{NN\to NN\rho} (mb)=\frac{0.24(\sqrt{s}-\sqrt{s_0})}{1.4+(\sqrt{s}-\sqrt{s_0})^2}
\end{equation}
with $\sqrt{s_0}=2.646$ GeV being the threshold of the reaction.
In view of the strong coupling of the $\rho$ to nuclear resonances
this course-grained parametrization has most probably large
systematic errors and presents a lower limit to the $\rho$
production. Other models like URQMD use a parametrization of the
resonance production which yield higher $\rho$ yields.
 For the $\pi$+N $\to$ N+$\rho$ data \cite{lutz}
we use the parametrization of \cite{pinomegarho}
\begin{equation}
\sigma_{\pi N\to \rho N} (mb)=\frac{1.5(\sqrt{s}-\sqrt{s_0})^{2.2}}{0.0018+(\sqrt{s}-\sqrt{s_0})^{3.5}}
\end{equation}
with $\sqrt{s_0}=1.708$ GeV.

Having a large width and therefore a short life time, the $\rho$
meson is an ideal particle to probe whether the nuclear environment
changes mesonic properties. If produced in hadronic matter the
majority of them decay in matter and therefore the dileptons carry
direct information on the in-medium properties. Theory predicts that
these properties are different from that of the free $\rho$. Whereas
there seems to be now consensus that the width of the $\rho$
increases if brought into a nuclear environment
\cite{pis,rap1,rapp}, the question of how the pole mass changes is
still debated. Based on QCD sum rule calculations, Hatsuda and Lee
\cite{hat} predicted a lowering of the $\rho$ mass in a nuclear
environment, a suggestion which has later been confirmed by Brown
and Rho \cite{bro1,bro2}. More recent  and more sophisticated
calculations leave, on the contrary,  the $\rho$ mass almost
unchanged \cite{rap1,kapu,rapp}. Experimentally the situation is also far
from being clear. In pA collisions \cite{om} at 12 GeV a decrease of
the mass ($m(\rho)/m(0) = 1 - 0.09 \rho /\rho_0$  - about half of
the value predicted by theory) and no increase of the width has been
reported. The dilepton data in In+In collisions at 158 AGeV
\cite{na60} are best described using the free $\rho$ pole mass but a
considerable broadening of the mass distribution. In contradiction
to the earlier theoretical expectations this broadening is almost symmetric
around the pole mass but recently it has been pointed out \cite{renk1} that 
the $\Phi$-functional approach may explain this symmetry.

Whether these experimental differences are
exclusively due to the different environments (cold nuclear matter
in pA reactions, an expanding meson dominated fireball after a
possible phase transition from a quark gluon plasma in AA
collisions) has not been fully explored yet. It is very difficult to
exploit this experimental information for heavy ion reactions at 2
AGeV where theory predicts that most of the $\rho$ mesons are  decay
products from nuclear resonances, especially of the $N^*(1520)$
resonance which has a branching ratio of $15-25\%$ into the $\rho$N
channel. For the present status of the theoretical spectral function
calculations for the $\rho$ meson we refer to \cite{mos,rap3}.

As for the $\omega$ meson the inconclusive situation of theory and experiment
suggest to employ for this exploratory study the free pole mass distribution
of the $\rho$:
\begin{equation}
\frac{dN}{dM}=\frac{m_{\rho}^2}
{(\frac{M^2-m_{\rho_2}^2}{m_{\rho}})^2+\Gamma_{\rho}^2}
\end{equation}
with
$m_{\rho}=0.775$ GeV/c$^2$, $m_{\rho_2}=0.761$ GeV/c$^2$,
$\Gamma_{\rho}=0.118$ GeV/c$^2$ \cite{sak} and the parametrized free cross sections. For the
branching ratio of the $\rho$ into dileptons we use $4.5\cdot 10^{-5}$.

\subsection{pn - bremsstrahlung}
In each np collisions real and virtual photons can be produced. The invariant mass 
distribution of the $e^+e^-$ pairs, the decay product of the virtual photon, 
is given by:
\begin{equation} \label{brehm}
\frac{dP(s,M)}{dM}=\frac{1}{3}\frac{\alpha^2}{\pi^2}\frac{1}{M}\frac{s-(m_p+m_n)^2}{e_{cm}^2}ln(\frac{q_{max}+q_{0max}}{M}-\frac{q_{max}}{q_{0max}})
\end{equation}
with
\begin{equation}
q_{0max}=\frac{s+M^2-(m_p+m_n)^2}{2\sqrt{s}}
\end{equation}
\begin{equation}
q_{max}=\sqrt{q_{0max}^2-M^2}.
\end{equation}
$\sqrt{s}$ is the np center of mass energy, $e_{cm}$ is the energy of
the incoming proton in the np center of mass system, $\alpha$ is the
electromagnetic coupling constant, $m_p$ and $m_n$ the masses of
proton and neutron, $q_{0max}$ the maximal dilepton energy and
$q_{max}$ the maximal dilepton momentum. The bremsstrahlung from pp
collisions is of quadrupole type and can be neglected as compared to
the dipole pn bremsstrahlung.

\subsection{$\Delta$ Dalitz decay}
It is not experimentally verified yet whether the Dalitz decay into
$e^+e^-$ of the $\Delta$ resonance exists but since it decays into a
photon it should also decay into a dilepton. The width of the
Dalitz-decay to dileptons of invariant mass $M$ is determined by 
QED \cite{Lautrup}:
\begin{equation}
{d\Gamma \over dM^2} = {\alpha \over 3\ \pi } {\Gamma _0(M^2)\over M^2}
\end{equation}
where
\begin{equation}
\Gamma _0(M^2) =
{\lambda ^{1/2}(M^2,m_N^2,m_\Delta ^2)\over 16\ \pi \ m_\Delta ^2} m_N [
2\ M_t(M^2) + M_l(M^2) ]
\end{equation}
is the total decay rate into a virtual photon with mass $M$ and
\begin{equation}
\lambda (x,y,z) = x^2 + y^2 + z^2 - 2 (xy + xz + yz).
\end{equation}

\medskip
\noindent
$M_t$ and $M_l$ depend on the form of the interaction. For the $\Delta $ decay
we take the $N\Delta \gamma $ vertex from \cite{ndeltagammavertex}. Using this
interaction we obtain the following matrix elements:
\begin{eqnarray}
M_l &=& (e\ f\ g)^2 {m_\Delta ^2\over 9\ m_N} M^2\ 4 (m_\Delta  - m_N - q_0)
          \nonumber \\
M_t &=& (e\ f\ g)^2 {m_\Delta ^2\over 9\ m_N} [q_o{}^2(5\ m_\Delta  - 3(q_0 +
m_N)) - M^2(m_\Delta  + m_N + q_0)]
\end{eqnarray}

with 

\begin{equation}
f=-1.5 \frac{m_\Delta+m_N}{m_N((m_N+m_\Delta)^2-M^2)},
\end{equation}

$q_0$ the energy of the dilepton pair in the $\Delta$ center of mass, $e$ the electric charge and $g = 2.72$ is the coupling constant fitted to the photonic decay width
$\Gamma_0(0) = 0.72$ MeV \cite{wwolf}.

\section{The C+C reaction at 2 and 1 AGeV}
For the simulation of the heavy ion reaction we use the 
IQMD program which has been described in section I.
The details of this program can be found in \cite{hartn}.

The presented results are impact parameter averaged and have been
corrected for the experimental mass resolution and acceptance with a
program provided by the HADES collaboration. We have neglected in
our calculation the reabsorption cross section of the $\eta$ mesons
which is of the order of 20 mb \cite{bha} in our kinematic domain
but of little importance for such a light system. We compare the
results of the standard set up, where free masses and widths as well
the most common extrapolations or theoretical predictions 
of unknown cross sections are used, with calculation in which
it is assumed that the particle properties change in the medium 
or in which other cross section parametrization are applied.

\begin{figure}[!ht]
\begin{center}
\includegraphics[width=7.0cm]{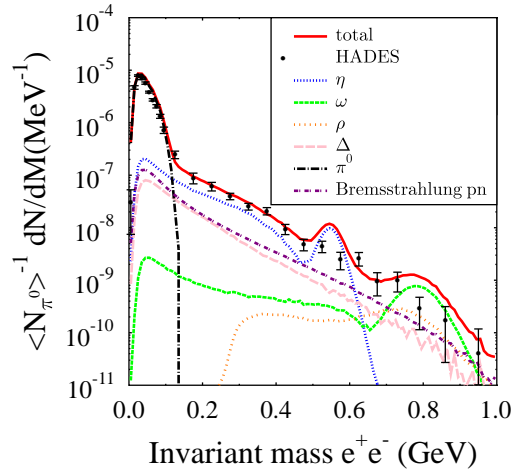}
\end{center}
\caption{(Color online) The invariant mass spectrum of the HADES collaboration as compared with IQMD simulations
for C+C at 2AGeV using $\sigma (np \to np\eta) = 2 \sigma(pp \to pp\eta)$ , $\sigma (np \to np\omega) = 5 \sigma (pp \to \omega),
M_{\omega} = M_{\omega}^0$ and the branching ratio ($\eta \to e^+e^-$) = 7.7 10$^{-5}$ (model A).}
\label{van1}
\end{figure}

\begin{figure}[!ht]
\begin{center}
\includegraphics[width=7.0cm]{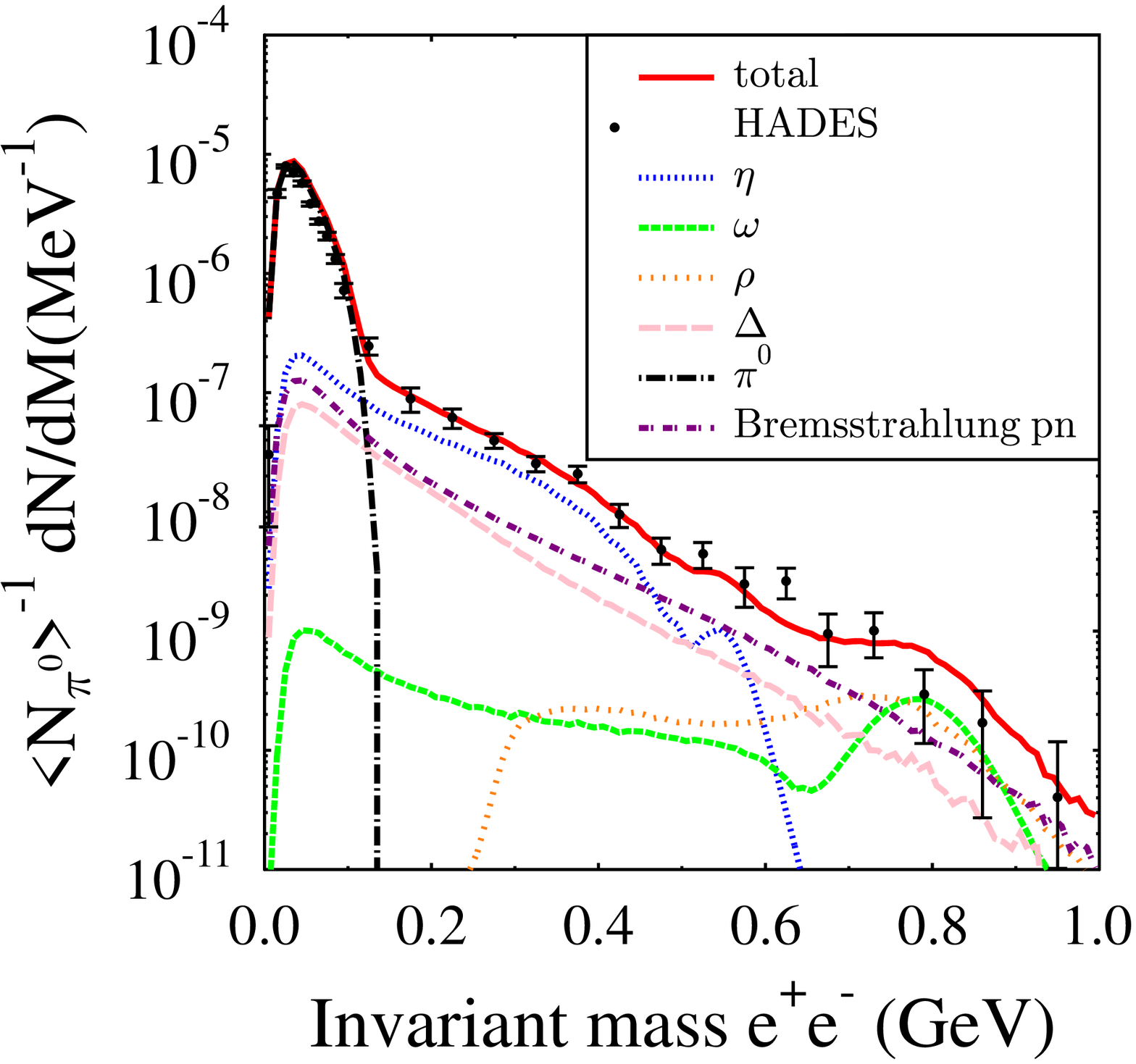}
\end{center}
\caption{(Color online) The invariant mass spectrum of the HADES collaboration as compared with IQMD simulations
for C+C at 2AGeV using $\sigma (np \to np\eta) = 2\sigma(pp \to pp\eta)$ , $\sigma (np \to np\omega) =  \sigma(pp \to pp\omega)$
$M_{\omega} = M_{\omega}^0$ and the branching ratio ($\eta \to e^+e^-$) = 7.7 10$^{-6}$ (model B).}
\label{van2}
\end{figure}

\begin{figure}[!ht]
\begin{center}
\includegraphics[width=7.0cm]{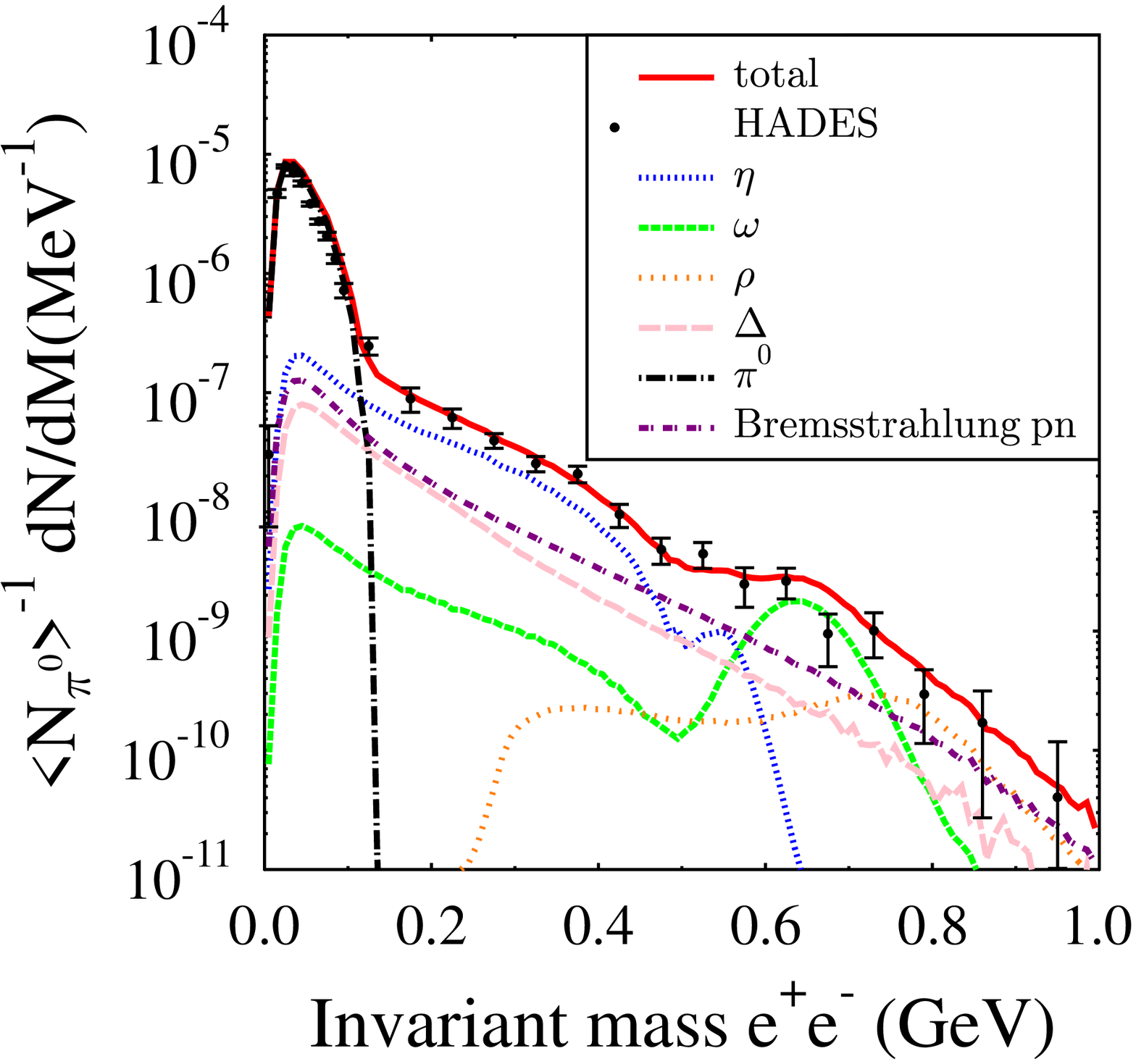}
\end{center}
\caption{(Color online) The invariant mass spectrum of the HADES collaboration as compared with IQMD simulations
for C+C at 2AGeV using $\sigma (np \to np\eta) = 2\sigma(pp \to pp\eta)$ , $\sigma (np \to np\omega) = 5 \sigma(pp \to pp\omega)$
$M_{\omega} = M_{\omega}^0(1-0.13\rho/\rho_0)$ and the branching ratio ($\eta \to e^+e^-$) = 7.7 10$^{-6}$ (model C).}
\label{van3}
\end{figure}

\begin{figure}[!ht]
\begin{center}
\includegraphics[width=7.0cm]{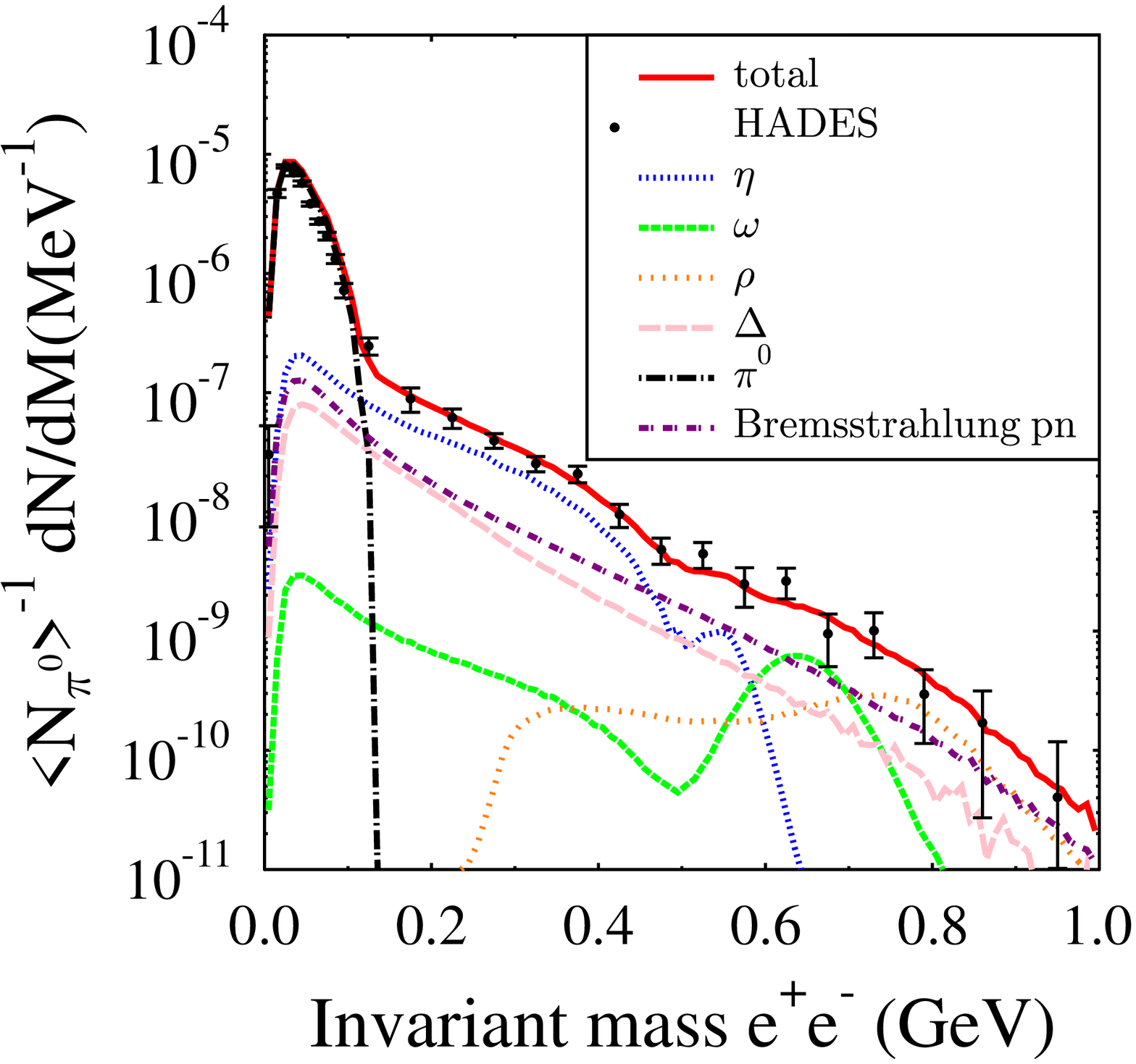}
\end{center}
\caption{(Color online) The invariant mass spectrum of the HADES collaboration as compared with IQMD simulations
for C+C at 2AGeV using $\sigma (np \to np\eta) = 2 \sigma(pp \to pp\eta)$ , $\sigma (np \to np\omega) = \sigma (pp \to pp\omega)$,
$M_{\omega} = M_{\omega}^0(1-0.13\rho/\rho_0)$ and the branching ratio ($\eta \to e^+e^-$) = 7.7 10$^{-6}$ (model E).}
\label{van5}
\end{figure}

Fig. \ref{van1} shows the result of the standard simulation set up: $\sigma (np \to
np \eta) = 2 \sigma(pp \to pp\eta)$ , $\sigma (np \to np\omega) = 5
\sigma(pp \to pp\omega)$, $M_{\omega} = M_{\omega}^0$ and the
branching ratio BR$_{\eta \to e^+e^-} = 7.7 \cdot 10^{-5}$. 
It is called standard because it uses standard literature values 
for the unknown physical input quantities. We see
first of all that with the resolution of the HADES experiment the
direct $\eta$ decay would yield a visible peak which is not present
in the data. Therefore the upper limit has to be lower than that
quoted by the Particle Data Group \cite{rev}. We see as well that the
simulations overpredict the yield in the region of the free $\omega$
mass. This confirms the result of the simulations with other
programs which have been published by the HADES collaboration
\cite{hades}. On the contrary, the simulations reproduce well the
mass region in which the lepton pairs are coming dominantly from the
$\eta$ decay and from pn bremsstrahlung. If the experimentally
unknown $\sigma (np \to {pn}\omega)$ equals $\sigma (pp \to
{pp}\omega)$ the yield in the $M_{\omega}$ mass region would be strongly
reduced and comes closer to experimental data as can be
seen in fig. \ref{van2}.  We obtain the same level of agreement with data if
we take $\sigma (np \to pn\omega)$ = $5\sigma (pp \to pp\omega)$ but
assume in addition that the mass of the $\omega$ decreases in the
medium according to eq. \ref{scal}, as indicated by the CBELSA/TAPS
results \cite{taps}. This can be seen in fig. \ref{van3}. The experimental
error bars are large, however, do not show much structure and the 
deviations are only a factor of two. Higher statistics data would certainly 
improve this situation. The
best agreement is obtained in simulations with $\sigma (np \to {pn}\omega)$ = $\sigma (pp \to
{pp}\omega)$ and an in medium $\omega$ mass, as seen in fig. \ref{van5}.

Therefore, without further informations on $\sigma (np \to np\omega)$ heavy ion
reactions will not reveal any robust information on in medium modifications
of the $\omega$ meson.

\begin{figure}[!ht]
\begin{center}
\includegraphics[width=7.0cm]{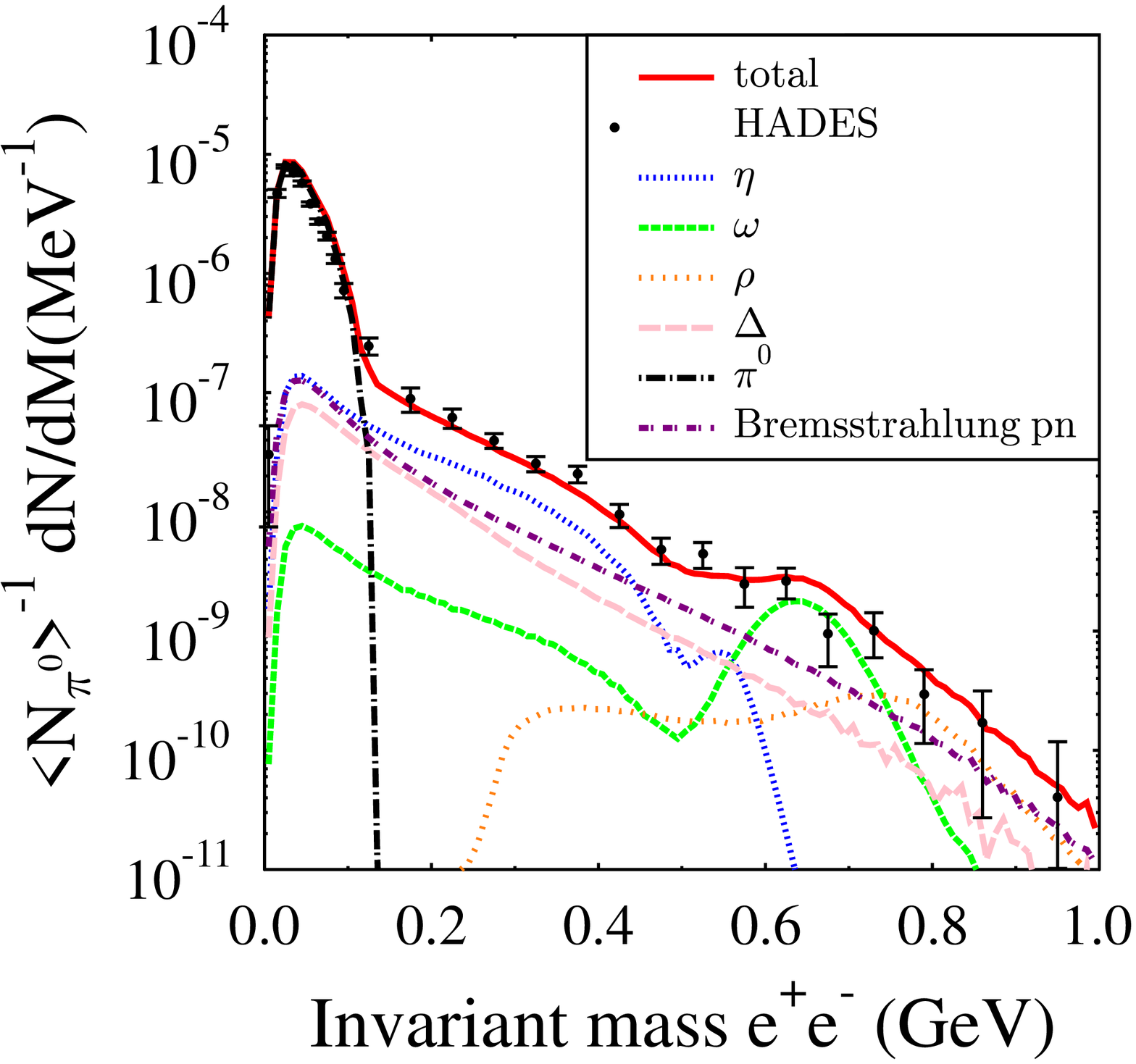}
\end{center}
\caption{(Color online) The invariant mass spectrum of the HADES collaboration as compared with IQMD simulations
for C+C at 2AGeV using $\sigma (np \to np\eta) = \sigma(pp \to pp\eta)$ , $\sigma (np \to np\omega) = 5 \sigma(pp \to pp\omega)$
$M_{\omega} = M_{\omega}^0(1-0.13\rho/\rho_0)$ and the branching ratio ($\eta \to e^+e^-$) = 7.7 10$^{-6}$ (model D).}
\label{van4}
\end{figure}

If we assume $\sigma (np \to np\eta) = \sigma(pp \to pp\eta)$ in the
region where no data on the $\sigma (np \to np\eta)$ cross section
is available (only 5\% of the $\eta$ are produced at an energy where
experimental information on this cross section is available), we 
underpredict slightly the yield in this mass region as seen in fig.
\ref{van4}. The experimental error bars are too large, however, 
in order to conclude more than that there are indications
that if the mass of the $\eta$ does not change in the medium 
the $\sigma (np \to np\eta)$ is larger 
than $\sigma(pp \to pp\eta)$ at excess energies above 100
MeV. As for the $\omega$ meson possible in medium changes of the
$\eta$ meson require a detailed study of  its production in the pn
channel.


\begin{figure}[!ht]
\begin{center}
\includegraphics[width=15.0cm]{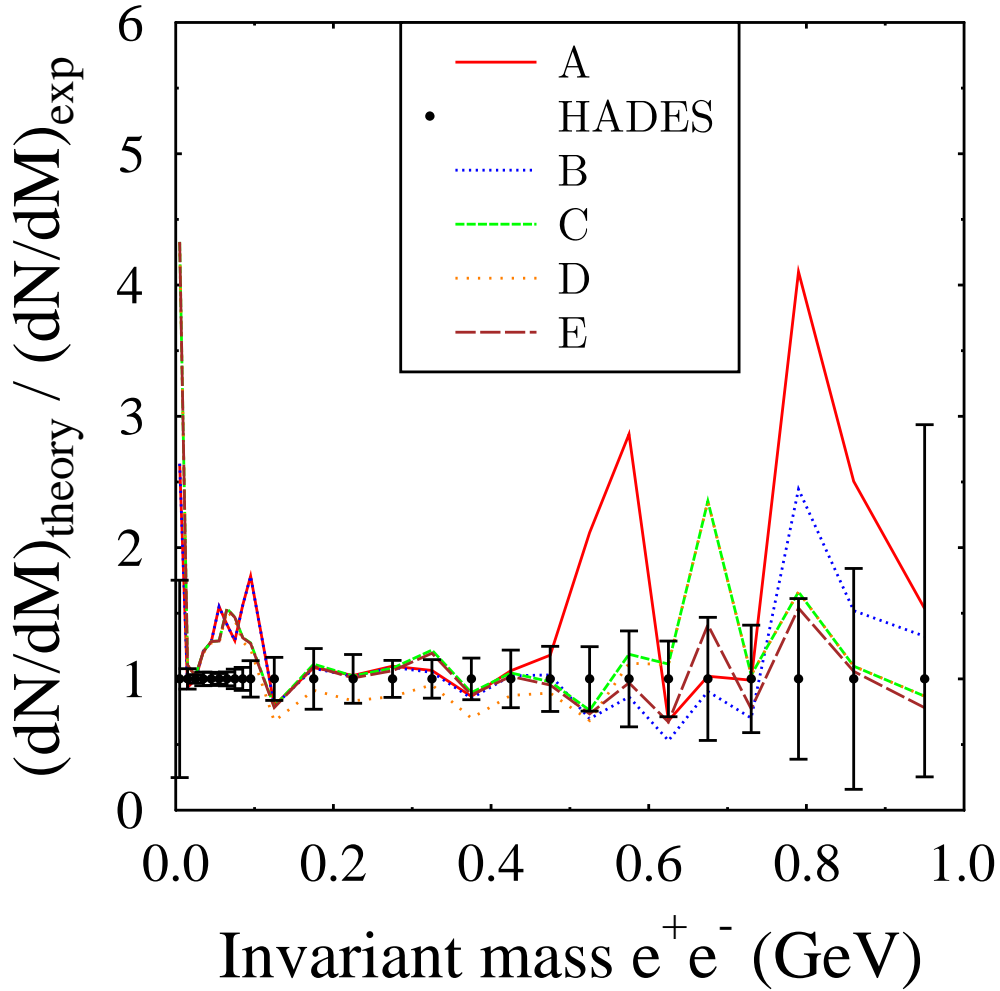}
\end{center}
\caption{(Color online) Invariant mass dependence of the ratio theory/experiment for the C+C 
reaction at 2 AGeV
for the different parametrization of unknown physical input quantities (see table III for details)}
\label{comp}
\end{figure}

Fig.\ref{comp} summarizes the study of the influence of the parametrization of unknown processes on the dilepton yield.
If one compares the results of the different scenarios of table III with the experimental results, we see that the standard
parameterizations of these input quantities (A) yield not results which are in agreement with data at invariant dilepton 
masses around 550 MeV
and in between 750 MeV and 950 MeV. The former difference suggests that the partial width for the disintegration of the 
$\eta$ into a dilepton pair is much smaller than the upper limit quoted by the Particle Data Group \cite{rev}. The latter
discrepancy contains the interesting physics as far as in medium particle properties are concerned. We see that 
even a reduced $\omega$ production cross section in the np channel (B) does not render the calculation
compatible with the data. Also the assumption that the mass of the $\omega$ changes in the medium but that it is produced
with the free cross section (C) overpredicts the experimental results because it shifts the surplus only to lower invariant 
masses.  Only the combination of a lower in medium mass and a reduction of the standard assumption on the cross section in 
the pn channel (E) yield results which are compatible with the experimental error bars. The scenario (D) demonstrates that 
the data are not sufficiently precise to allow for robust conclusions on the $np \rightarrow np\eta$ channel. 
A variation of a factor of two gives results which are both compatible with experiment.
 
Thus the C+C data at 2 AGeV show interesting new physics which is not compatible with the input of
state of the art transport codes. Unfortunately without further information on the elementary cross sections
with a neutron in the entrance channel it will not be possible to identify the origin of this discrepancy
because a modification of the mass of the mesons in the medium yields the same effect as a change of the
(experimentally unknown) cross section in the np channel.

\begin{figure}[!ht]
\begin{center}
\includegraphics[width=7.0cm]{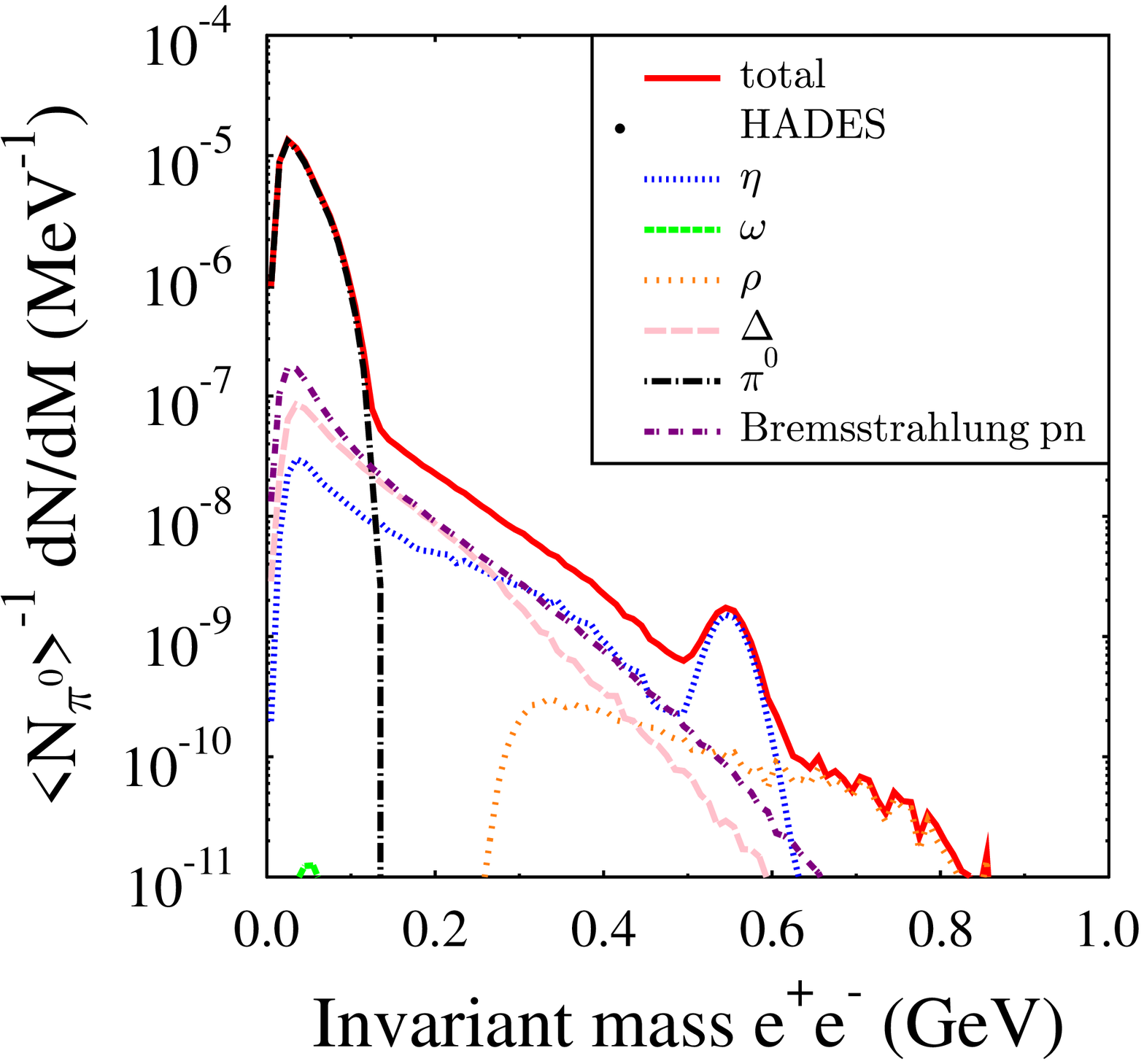}
\end{center}
\caption{(Color online) The invariant mass spectrum of IQMD simulations for C+C at
1AGeV using $\sigma (np \to np\eta) = 2 \sigma(pp \to pp\eta)$ ,
$\sigma (np \to np\omega) = 5 \sigma (pp \to pp\omega)$,
 $M_{\omega} = M_{\omega}^0$ and the branching ratio ($\eta \to e^+e^-$) = 7.7 10$^{-5}$.}
\label{van11}
\end{figure}

\begin{figure}[!ht]
\begin{center}
\includegraphics[width=7.0cm]{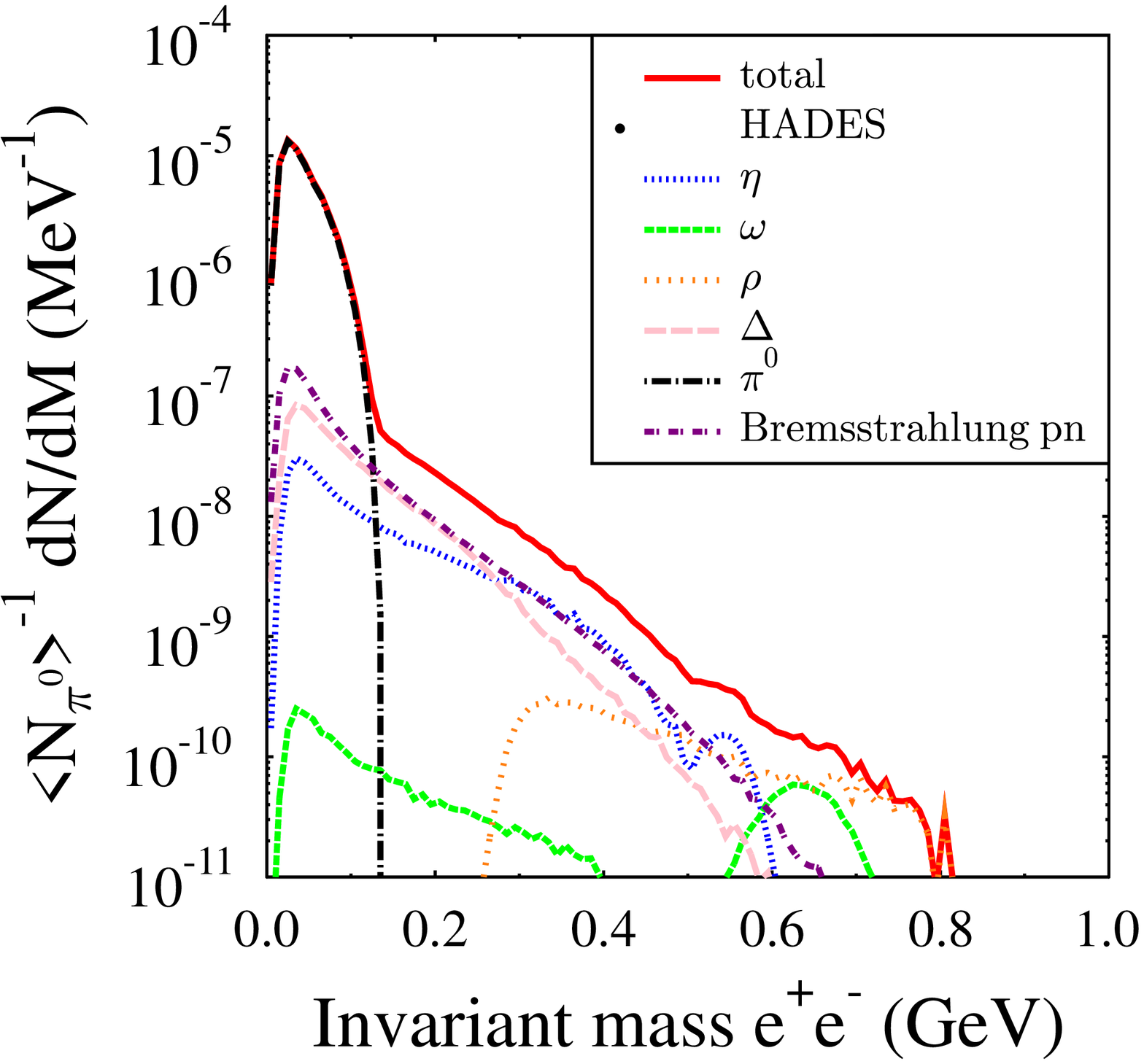}
\end{center}
\caption{(Color online) The invariant mass spectrum of IQMD simulations for C+C at
1AGeV using $\sigma (np \to np\eta) = 2 \sigma(pp \to pp\eta)$ ,
$\sigma (np \to np\omega) = 5 \sigma (pp \to pp\omega)$, $M_{\omega} =
M_{\omega}^0(1-0.13\rho/\rho_0)$ and the branching ratio ($\eta \to
e^+e^-$) = 7.7 10$^{-6}$.} \label{van13}
\end{figure}

Lowering the energy to 1 AGeV  the importance of the different
channels changes and a comparison between the 2 and 1 AGeV data will
elucidate part of the physics. Because the experimental data are
divided by the number of $\pi^0$ the spectra for the pions change
only little due to the acceptance corrections. The same is true for
the $\Delta$ Dalitz decay. The yield of $e^+e^-$ pairs from $\eta$
Dalitz decay and bremsstrahlung are lower, on the contrary, and the
$\omega$ production is practically absent due to the lack of energy
(even if one takes into account that the Fermi momentum may create a
larger $\sqrt{s}$ value than in NN collisions at the same beam
energy). Fig. \ref{van11} displays our filtered and acceptance
corrected results. In the intermediate mass region the $\Delta$
Dalitz decay and bremsstrahlung have gained importance and are of
the same order of magnitude. Dilepton pairs from $\eta$ Dalitz decay
are less frequent and are not dominant anymore in the intermediate
mass region. At this energy about 35\% of the $\eta$ come from a
$\sqrt{s}$ region where the np production cross section is known. So
the uncertainty in this channel is reduced but still present. In the
standard set up of the simulations the dilepton invariant mass
spectrum at intermediate masses has always a
strong component of the bremsstrahlung which gives about 50\% of the
yield. At invariant masses of around 200 MeV the $\Delta$ Dalitz
decay contributes the other 50\% - if it exists. The data at 1 AGeV
should therefore allow do define an upper limit of the  $\Delta$
Dalitz decay. At higher invariant masses it is the $\eta$ decay
which contributes the other 50\%. If we assume that $\sigma (np \to
np\eta) = \sigma(pp \to pp\eta)$  the $\eta$ yield becomes that low
that its influence on the spectrum is hardly visible.
If we lower the in medium $\omega$ mass we see a larger $\omega$
production cross section but it remains a small contribution to the
total yield, as seen in fig. \ref{van13}.

\begin{figure}[!ht]
\begin{center}
\includegraphics[width=7.0cm]{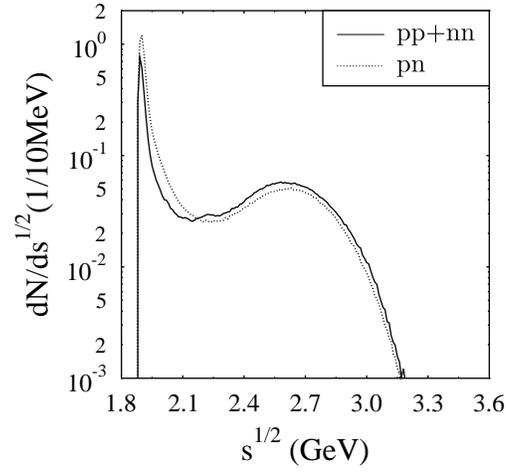}
\end{center}
\caption{Distribution of $\sqrt{s}$ of NN collisions in a C+C
reaction at 2 AGeV} \label{csdiffpp2GeV}
\end{figure}

It is interesting to see in detail the differences between
elementary collisions at $\sqrt{s} = 2.697$ GeV and heavy ion
collisions at the same nominal energy which show the large
$\sqrt{s}$ distribution of the NN collisions displayed in fig.
\ref{csdiffpp2GeV}. There we see two peaks. The high energy peak is
due to collisions between projectile and target nucleons, whereas
the low energy peak is due to collisions among either projectile or
target nucleons. The latter collisions contribute only to the
bremsstrahlung and to the $\pi^0$ part of the dilepton spectrum.
Due to rescattering the maximum of the distribution of the primary
collisions is shifted toward a lower $\sqrt{s}$ value. The consequences of the
broad $\sqrt{s}$ distribution on the $\eta$ and $\omega$ production
as compared to elementary collisions at the nominal energy are summarized in table
\ref{multiplicite2GeV}. The first line shows the average $\sqrt{s}$
value of all collisions above threshold. For the $\eta$ this value is
slightly below, for the $\omega$ - due to the larger threshold -
slightly above the value for elementary collisions. Also the average
number of collisions in C+C reactions depends on the particle type
as seen in the second line. For the $\eta$ production we find 4.65
collisions above threshold for the $\omega$ production  2.32. For
the standard scenario  ($m_\omega=m_\omega^0$, BR($\eta \to
e^+e^-)=7.7.10^{-5}, \sigma(pn \to pn\eta) = 2\sigma (pp \to
pp\eta)$ and $\sigma(pn \to pn\omega)=5\sigma(pp \to pp\omega)$), we
display in the third and the fourth line the average production
cross section in np and pp collisions in the heavy ion reaction as
compared to the elementary reaction. For the $\eta$ the average
$\sigma(pp \to pp\eta)$ and $\sigma(pn \to pn\eta)$ are lower in CC
collisions than in elementary ones. This decrease has two origins:
firstly the lower $<\sqrt{s}_ {coll>threshold}>$  and secondly the
form of the $\eta$ production cross section which has a maximum at
around $\sqrt{s}$=2.697 GeV and stays almost constant at higher
energies. For the $\omega$ meson the situation is completely
different. The elementary cross section increases with energy for
all relevant energies and the average $\sqrt{s}$ value in C+C is
larger than that in elementary collisions. Therefore np as well as pp
collisions in the heavy ion reaction produce more $\omega$ mesons
than elementary collisions at the same nominal energy.
Consequently the enhancement factor of $\eta$ and $\omega$ mesons in
heavy ion collisions is very different.

\begin{figure}[!ht]
\begin{center}
\includegraphics[width=7.0cm]{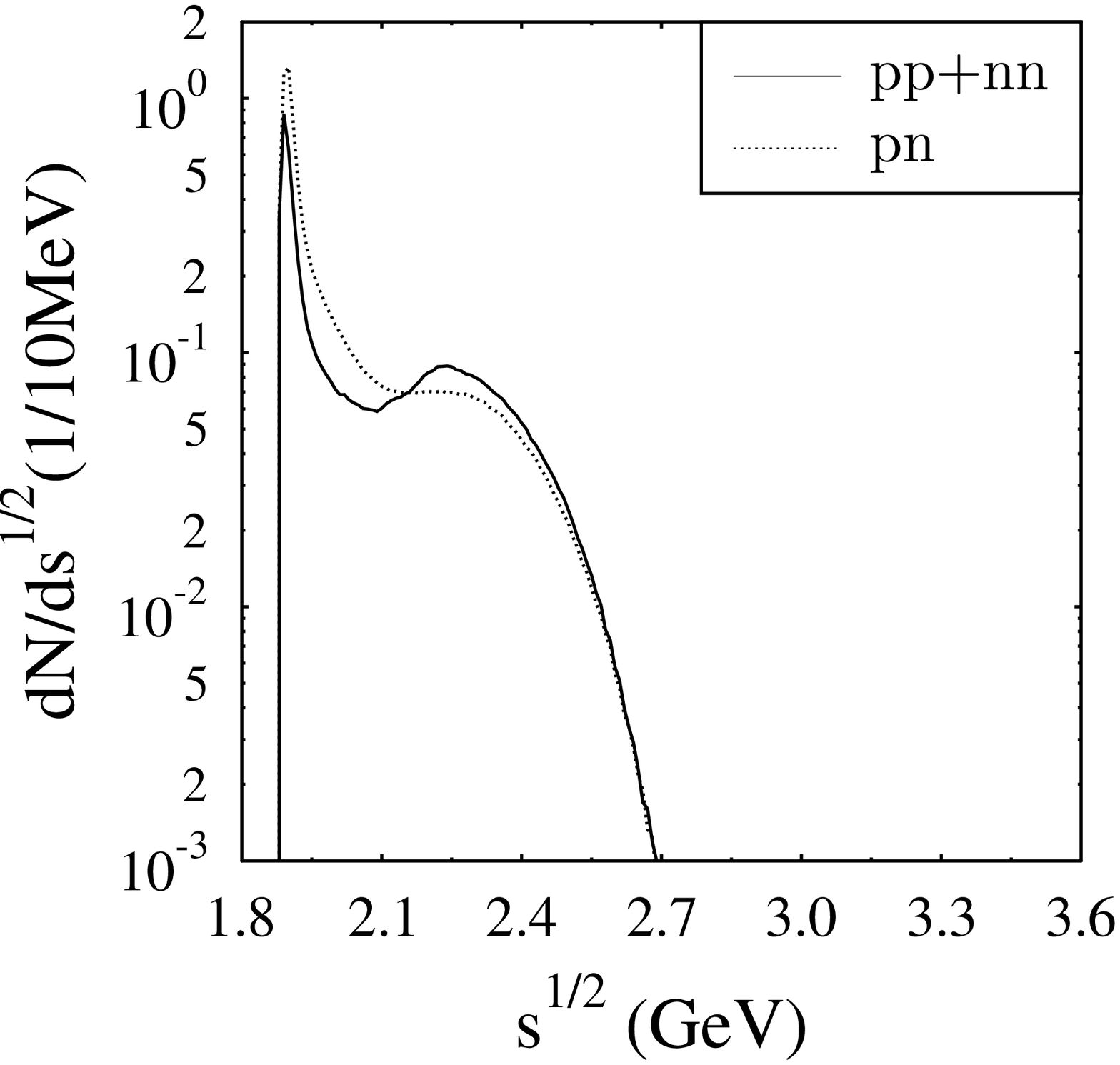}
\end{center}
\caption{Distribution of $\sqrt{s}$ of NN collisions in a C+C reaction at 1 AGeV}
\label{csdiffpp1GeV}
\end{figure}

For the  1 AGeV reaction (table \ref{multiplicite1GeV}) the situation
is very different. The $\sqrt{s}$ distribution of the collision in
C+C is displayed in fig. \ref{csdiffpp1GeV}. In elementary NN 
collisions at the same energy neither $\omega$ nor $\eta$ mesons can be produced (
$\sqrt{s}_{threshold \textrm{ }\omega}$ = 2.659 GeV and $\sqrt{s}_{threshold \textrm{ } \eta}$ = 2.424 GeV). 
However, with the Fermi momentum, in C+C
collisions subthreshold $\omega$ and $\eta$ production is possible.
Due to the larger threshold $\omega$ production is suppressed with
respect to the $\eta$ production. The production cross section at
this energy tests the Fermi motion in the simulations which is not
easy to model in semi-classical simulation codes. Therefore
systematic errors reduce the predictive power for the meson
production at this energy, but the analysis of the subthreshold kaon
production shows that in between a factor of two the
results are certainly trustworthy.

In summary we have shown that the dilepton spectrum measured by the
HADES collaboration in the reaction 2 AGeV C+C at invariant masses
above 600 MeV is not compatible with the standard scenario of
simulation programs which uses free cross sections and free meson
masses. Introducing a medium modification of the $\omega$ mass and
lowering the unknown $pn \to pn\omega$ cross section brings the
calculation in agreement with data. The extrapolation from
elementary cross section at the same nominal energy to heavy ion
reactions is all but trivial. It depends on the threshold and on the
energy dependence of the cross section. Before the elementary
production cross sections in pn reactions are not determined and
before the cross sections for baryonic resonances are not better
known heavy ion data do not provide the desired information on
possible in medium modification of the meson properties.


\begin{table}[!ht]
\begin{center}
\begin{tabular}{|c|c|c|c|c|c|c|}
\hline
Particle & \multicolumn{3}{|c|}{$\eta$} & \multicolumn{3}{|c|}{$\omega$}   \\
\hline
Collision & p+p & p+n & C+C & p+p & p+n & C+C \\
\hline
$<\sqrt{s}_{coll>threshold}>$ & 2.697 & 2.697 & 2.677 & 2.697 & 2.697 & 2.811  \\
\hline
$\langle N_{coll>threshold}\rangle$ & 1 & 1 & 4.65 & 1 & 1 & 2.32  \\
\hline
$\langle \sigma_{prod}\rangle_{C+C}$ ($\mu b$) & 115 & 304 & 203 & 23.8 & 115 & 66.5  \\
\hline
$ \sigma_{prod}$ ($\mu b$) & 175 & 359 &  & 4.90 & 24.51 &   \\
\hline
Multiplicity & 3.89.10$^{-3}$ & 8.37.10$^{-3}$ & 2.25.10$^{-2}$ & 2.36.10$^{-4}$ & 5.57.10$^{-4}$ & 3.6.10$^{-3}$ \\
\hline
\end{tabular}
\end{center}
\caption{Comparison of the average number of collisions above
threshold, of the average production cross section per NN collision in CC
collisions, the cross section in elementary pp and pn reactions 
and the multiplicity in pp, pn and CC collisions at a beam energy
of 2 AGeV for $\eta$ and $\omega$ mesons.} 
\label{multiplicite2GeV}
\end{table}
\begin{table}[!ht]
\begin{center}
\begin{tabular}{|c|c|c|c|c|c|c|}
\hline
Particle & \multicolumn{3}{|c|}{$\eta$} & \multicolumn{3}{|c|}{$\omega$}   \\
\hline
Collision & p+p & p+n & C+C & p+p & p+n & C+C \\
\hline
$<\sqrt{s}_{coll>threshold}>$ & x & x & 2.498 & x & x & 2.711  \\
\hline
$ N_{coll>threshold} $ & x & x & 0.812 & x & x & 0.0176  \\
\hline
$\langle \sigma_{prod}\rangle_{C+C}$ ($\mu b$) & 30.2 & 146.3 & 84.7 & 7.11 & 38.3 & 22.66  \\
\hline
$\langle \sigma_{prod}\rangle$ ($\mu b$) & 0 & 0 & & 0 & 0 & \\
\hline
Multiplicity & 0 & 0 & 1.61.10$^{-3}$ & 0 & 0 & 8.34.10$^{-6}$ \\
\hline
\end{tabular}
\end{center}
\caption{Comparison of the average number of collisions above
threshold, of the average production cross section per NN collision in CC
collisions, the cross section in elementary pp and pn reactions 
and the multiplicity in pp, pn and CC collisions at a beam energy
of 1 AGeV for $\eta$ and $\omega$ mesons.} 
\label{multiplicite1GeV}
\end{table}

\begin{table}[!ht]
\begin{center}
\begin{tabular}{|c|c|c|c|c|}
\hline
Model & BR$_{\eta\to e^+ e^-}$ & $m_{\omega}$ & $\frac{\sigma_{pn\to pn\omega}}{\sigma_{pp\to pp\omega}}$ & $\frac{\sigma_{pn\to pn\eta}}{\sigma_{pp\to pp\eta}}$   \\
\hline
A & 7.7 10$^-5$ & vacuum & 5 & 2  \\
\hline
B & 7.7 10$^-6$ & vacuum & 1 & 2  \\
\hline
C & 7.7 10$^-6$ & in-medium modification & 5 & 2  \\
\hline
D & 7.7 10$^-6$ & in-medium modification & 5 & 1  \\
\hline
E & 7.7 10$^-6$ & in-medium modification & 1 & 2  \\
\hline
\end{tabular}
\end{center}
\caption{Definition of the various parameters for the IQMD simulations} 
\label{modelsparam}
\end{table}

{\it Acknowledgment}: We acknowledge valuable discussions with M.
Bleicher, R. Holzmann, W. K\"uhn and J. Stroth and thank the HADES
collaboration for providing us with the filter routines. G. Wolf
acknowledge partial support by the grants T48833 and T47347.

\end{document}